\documentclass[11pt]{article}
\usepackage{natbib}
\usepackage[a4paper,top=2.5cm,bottom=2.5cm,left=3cm,right=3cm]{geometry}
\usepackage{longtable}
\usepackage{graphicx}
\usepackage{lscape}
\usepackage{graphics}
\usepackage{makecell}
\usepackage{environ}

\NewEnviron{smaller}{%
\scalebox{0.7}{\BODY}}

\begin{document}

\begin{centering}
\textbf{\large Jet-induced feedback in the [O~III] lines of early evolution stage active galactic nuclei} \\
\vskip 10pt
\small Marco Berton$^{a,b,}$\footnote{email: marco.berton@utu.fi} \& Emilia J\"arvel\"a$^{c,}$\footnote{email:ejarvela@sciops.esa.int} \\
\vskip 5pt
\footnotesize $^{a}$ Finnish Centre for Astronomy with ESO - University of Turku, Vesilinnantie 5, FIN-20500, Turku (Finland)\\
\footnotesize $^{b}$ Aalto University Metsähovi Radio Observatory, Metsähovintie 114, FIN-02450, Kylmälä (Finland)  \\
\footnotesize $^{c}$ European Space Agency (ESA), European Space Astronomy Centre (ESAC), Camino Bajo del Castillo s/n, 28692 Villanueva de la Ca\~nada, Madrid, Spain\\
\end{centering}

\begin{abstract}
It is well known that active galactic nuclei (AGN) show various forms of interaction with their host galaxy, in a number of phenomena generally called AGN feedback. In particular, the relativistic plasma jets launched by a fraction of AGN can strongly affect their environment. We present here a study of the [O~III] $\lambda\lambda$4959,5007 lines in a diverse sample of early evolution stage AGN -- specifically narrow-line Seyfert 1 galaxies. Radio imaging observations of all of the sources enable a division to jetted and non-jetted sources, and exploiting this we show that the ionized gas properties are significantly influenced by the presence of the jets, as we often find the [O~III] lines (blue-)shifted with respect to their restframe wavelength. We also show how the radio morphology and the radio spectral index do not seem to play a role in the origin of the [O~III] shifts, thus suggesting that the source inclination is not relevant to the lines displacement. We do not find a strong relation between the [O~III] line properties and the bolometric luminosity, suggesting that within our sample radiatively driven outflows do not seem to have a significant contribution to the [O~III] line kinematics. We finally suggest that [O~III] shifts may be a good proxy to identify the presence of relativistic jets. Additional studies, especially with integral-field spectroscopy, will provide a deeper insight into the relation between jets and their environment in early evolution stage AGN.
\end{abstract}

\vskip 5pt
\begin{centering}
\end{centering}

\newcommand{\kms}{km s$^{-1}$}
\newcommand{\ergs}{erg s$^{-1}$}
\newcommand{\Hb}{H$\beta$}
\section{Introduction}
Active galactic nuclei (AGN) play an essential role in the evolution of galaxies across cosmic time \citep{Taylor15}, as the central black hole (BH) and its host galaxy undergo co-evolution \citep{Kormendy13}. The nucleus can experience multiple activity phases throughout its life cycle, and during every phase there is interplay between the AGN and its host galaxy. The latter regulates the gas supply to the nucleus, enabling or hampering the triggering of the AGN, and inducing changes to its activity state. On the other hand, the AGN also affects its environment. The radiation produced by the process of accretion onto the BH can heat or ionize the gas up to several kiloparsecs (kpc) away \citep[e.g.,][]{Schmitt01, Congiu17b, Venturi18}, and by means of outflows or relativistic jets it can suppress or in some cases enhance \citep{Fragile17} the star formation activity across the host galaxy \citep{Morganti17}. All the examples of mutual interaction between an AGN and its host galaxy are collectively known as AGN feedback. \\
Feedback is still a relatively poorly understood aspect of galaxy evolution, and it is particularly interesting to study if and how this phenomenon occurs in AGN in early phases of their life cycle, that is when they are "young". Carrying out such research at high redshift, when all sources were young, is not an easy task. An alternative approach is to focus on sources that are the low redshift analogs to the first quasars. Narrow-line Seyfert 1 (NLS1) galaxies are in this sense ideal laboratories to study the early phases of AGN feedback since, like high-redshift quasars, they are characterized by fast growing black holes with high Eddington ratios, and also by high metallicity, especially prominent Fe II multiplets \citep{Mathur00}. \\
NLS1s are mostly non-jetted sources, but some of them do actually harbor powerful relativistic jets \citep{Komossa06, Yuan08, Abdo09a, Foschini11, Foschini15}. Typically hosted by spiral galaxies \citep{Krongold01, Crenshaw03, Deo06, Ohta07, Anton08, Orbandexivry11, Mathur12, Kotilainen16, Olguiniglesias17, Jarvela18, Berton19a, Olguiniglesias20}, due to their low-mass BHs and high Eddington ratios NLS1s are believed to be the progenitors of other AGN classes, such as broad-line Seyfert 1 galaxies and flat-spectrum radio quasars \citep{Grupe00, Mathur00, Sulentic00, Mathur01, Peterson11, Berton16c, Paliya19a, Berton20a, Foschini20}. Since their classification is exclusively based on the width of the H$\beta$ line \citep{Osterbrock85}, NLS1s can currently be found only up to z$\sim$1 \citep{Rakshit17a}, although unclassified objects with similar properties are present also at higher redshift \citep{Sulentic02, Marziani18b}. A feature which is often observed in the optical spectra of this class of AGN is a shift, usually toward lower wavelength, in the emission lines of [O~III]$\lambda\lambda$4959, 5007, corresponding to a velocity $|v| > 150$ \kms\ \citep{Komossa08}. Such sources, in the literature, are called blue outliers \citep{Zamanov02, Marziani03, Komossa08}. They are mainly found among NLS1s but, more in general, they are observed almost exclusively in population A sources on the quasar main sequence (MS). The MS is the locus on the plane defined by the flux ratio between the Fe~II multiplets and H$\beta$, known as R4570, and the FWHM(H$\beta$), where all type 1 AGN lie. This sequence was identified by means of principal component analysis, and it seems to be mainly driven by the Eddington ratio \citep{Boroson92}. However, some other factors seem to play a role in it, such as metallicity, and the inclination compared to the line of sight \citep{Shen14, Panda19}. Sources on the MS can be divided into two main populations according to the full width at half maximum (FWHM) of their H$\beta$ line. Population A sources are characterized by FWHM $<$4000 \kms, and they often exhibit prominent Fe~II emission \citep{Sulentic00, Marziani01, Sulentic02, Marziani03, Marziani18a, Czerny18}.
The origin of blue outliers is still unclear \citep{Marziani16, Komossa18a, Ganci19}, and it could be connected to the high Eddington ratio observed in population A sources and NLS1s \citep{Boroson02, Grupe04, Grupe10, Xu12}. However, their presence is often associated with strong radio emission indicative of relativistic jets \citep[][hereafter B16]{Komossa18a, Berton16b}. If this is true, blue outliers may be a remarkable example of feedback between kinematically young relativistic jets, as those present in NLS1s, and the ionized gas of their host galaxy. \\
In this work, we analyzed the properties of the [O~III] lines, with particular attention to blue outliers, in a sample of NLS1s whose radio morphology has been observed with the Karl G. Jansky Very Large Array (JVLA), to investigate their relation with radio emission. In Section~2 we describe our sample selection and the data analysis process, in Section~3 we present our results, in Section~4 we discuss our findings, and in Section~5 we present a brief summary of this work. Throughout the paper, we adopt the standard $\Lambda$CDM cosmology, with a Hubble constant H$_0 = 70$ \kms\ Mpc$^{-1}$, and $\Omega_\Lambda = 0.73$ \citep{Komatsu11}. 

\section{Sample selection and data analysis}

\begin{table}[t]
\caption{Observational details of the spectra observed with the Asiago 1.22m telescope. (1) Short name; (2) exposure time (seconds); (3) resolution; (4) observing date. The asterisk indicates that the spectra were observed on five separate dates, 2013-12-06, 2014-09-21, 2015-08-12, 2015-11-15, 2018-07-07, and later combined \citep{Tripathi20}. This operation is possible because the interval between these spectra is shorter than the typical scale of the NLR.}
\centering
\begin{tabular}{lccc}
\hline
Short name & Exp. Time & $R$ & Date \\
\hline
J0006+2012	& 25500 & 700 & 2013-2018* \\
J0347+0105	& 4800	& 700 & 2015-11-16 \\
J0629$-$0545 & 6000 & 700 & 2015-12-10 \\
J2242+2943 & 18000 & 700 & 2015-08-06 \\
\hline
\end{tabular}
\label{tab:obs}
\end{table}

Our sample includes the 74 NLS1s that were observed with the JVLA at 5~GHz in its most extended A configuration (longest baseline 36 km, resolution 0.5") in the survey described in \citet[][hereafter B18]{Berton18a}. We also decided to include into our sample seven additional NLS1s studied in the same JVLA configuration. These peculiar objects show mJy-level or lower flux density at 5~GHz, but they were observed flaring at Jy-level at 37~GHz, indicating that they harbor relativistic jets \citep{Lahteenmaki18}. The reason for the weak radio emission at low frequencies may be a form of absorption, either free-free or synchrotron self-absorption \citep{Berton20b}. 

We were able to retrieve an optical spectrum for 80 out of 81 sources (all but J0100-0200). In another object, J1348+2622, the [O~III] lines are not visible in the optical spectrum due to its high redshift ($z$ = 0.917), and it was therefore excluded from this analysis. For 62 of these NLS1s, a measurement of the [O~III] line properties was already available in B16. For the remaining 17 objects, 13 spectra were obtained from the Sloan Digital Sky Survey (SDSS) Data Release 16 (DR16) \citep{Blanton17}, while the remaining four were observed with the 1.22m telescope of the Asiago Astrophysical Observatory. The details of these new observations are reported in Table~\ref{tab:obs}. The measurement process we used is the same as the one described in detail in B16. In the following we provide a brief summary. \\
The first challenge was obtaining a precise estimate of the redshift of each source. Ideally, stellar absorption lines would be the best proxy for the recessional velocity of a galaxy, but in type 1 AGN they are typically not visible. Emission lines are easily detectable, but not all of them are reliable redshift indicators. Broad lines and high-ionization lines often originate from outflows and appear shifted with respect to the galaxy frame, so the ideal solution is to use low-ionization lines, such as [S~II]$\lambda\lambda$6717, 6731, [O~I] $\lambda$6300, or [O~II]$\lambda$3727 \citep{Komossa08, Sulentic08}. It is worth noting that the [O~II] line is an unresolved doublet in low-resolution spectra, therefore the other lines, if available, grant a better estimate of the redshift. However, due to the redshift, [O~II] is often the only available reference. We measured the wavelength of the reference line by using a fitting procedure in \texttt{Python}. In this procedure there are two sources of systematic error. The first one is due to wavelength calibration errors. For SDSS its value is 2 \kms\ \citep{Abazajian09}, while for the other spectra it is of the order of 20 \kms\ (B16). The other error is produced by the fit of the lines, and we computed it using a Monte Carlo method. We added a Gaussian noise, whose amplitude was proportional to the noise in the AGN continuum measured around $\lambda$5100, to the line. We then repeated the fitting procedure 100 times with different noise, getting a median value of redshift and a standard deviation. On average, the latter is rather small, $\sim$2 \kms\, as it is obtained by a simple fit with one Gaussian of a relatively strong emission line. \\
\begin{figure}[!t]
\centering
\includegraphics[width=0.6\hsize]{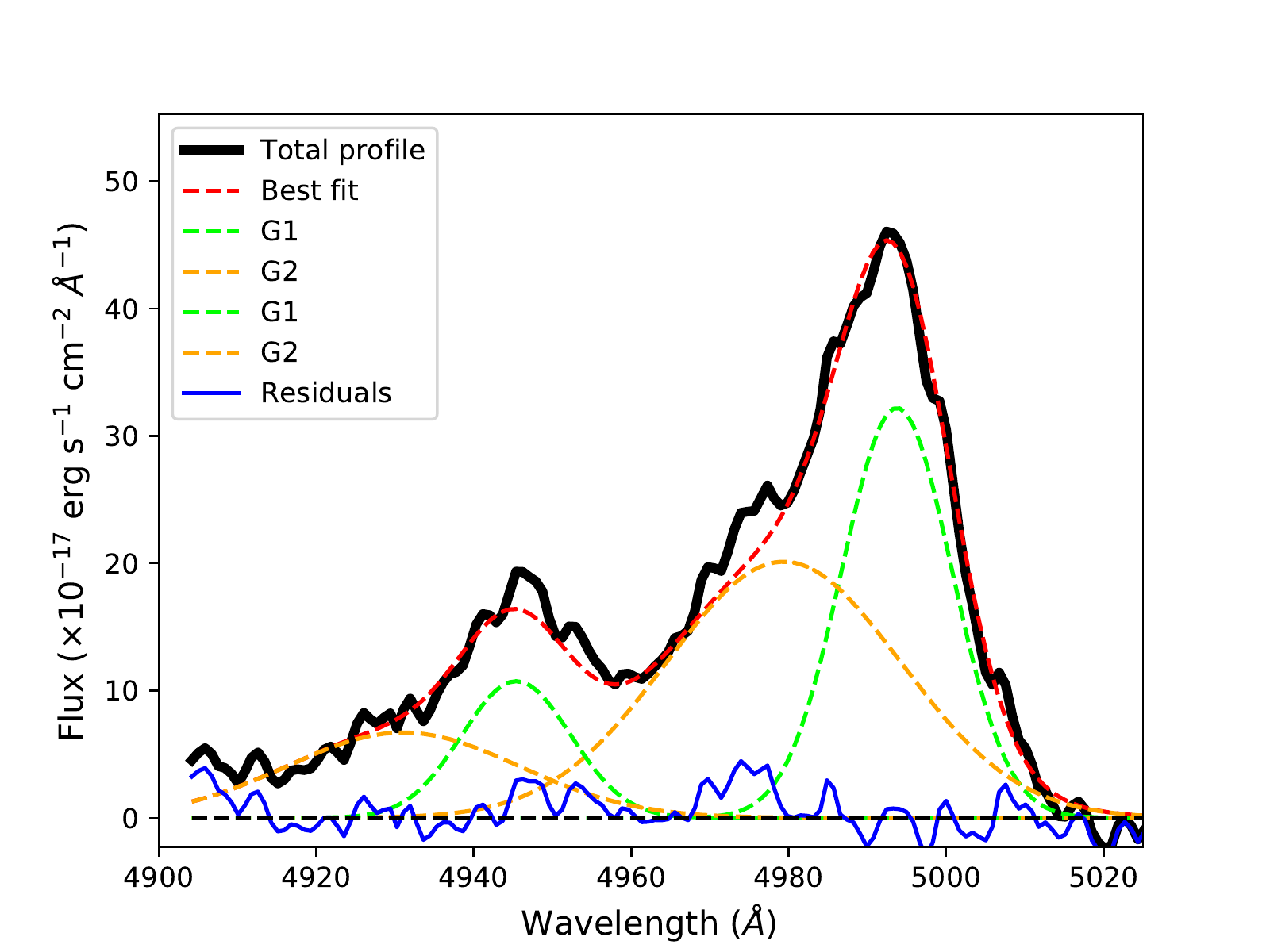}
\caption{[O~III] line fitting of the source J1114+3241. The black solid line represents the restframe spectrum after continuum and Fe~II subtraction, the green dashed line represents the [O III] core components, the orange dashed line represents the wing components, and the best fit is shown as the red dashed line. The blue solid line shows the residuals, and the black horizontal dashed line corresponds to a zero-level flux.}
\label{fig:J1114}
\end{figure}   
After moving the spectra to restframe, we subtracted the AGN continuum using IRAF and, when necessary, we removed the Fe~II multiplets using an online software\footnote{http://servo.aob.rs/FeII\_AGN/}\citep{Kovacevic10, Shapovalova12}. At this point, we decomposed each of the [O~III] lines using two Gaussian components, one to reproduce the line core, and one for the blue-shifted wing of the line. The line core component originates from the ionized gas of the narrow-line region (NLR). Its kinematics is dominated by the gravitational potential of the bulge, and therefore it usually has the same recessional velocity as the host galaxy. The wing component, instead, probably comes from a turbulent outflow in the inner part of the NLR. Since in type 1 AGN we preferentially have a pole-on view of the central engine, the outflow appears as a blue-shifted (approaching) component in the line with a FWHM larger than that of the core due to its turbulence. Our aim was obtaining the position of the [O~III] core component to derive the velocity of the ionized gas with respect to the galaxy frame, the position of the line wing to measure the velocity of the outflow, and the FWHM of both components to derive the gas velocity dispersion. To reduce the number of free parameters in the fitting procedure, we fixed the ratio between each component of the $\lambda$5007 and $\lambda$4959 lines to its theoretical ratio of 3.0 \citep{Dimitrijevic07}. Furthermore, we forced the FWHM of the line core, the FWHM of the wing, and the wing velocity, expressed in \kms, to be the same in both lines. The errors were computed adopting the previously described Monte Carlo procedure. In conclusion, for each line we measured four parameters: the core shift with respect to the laboratory wavelength of [O~III] (5006.843\AA), the line core FWHM, the wing velocity with respect to the core component, and the wing FWHM. An example of a fit is shown in Fig.~\ref{fig:J1114}. We successfully reproduced the [O~III] lines in 76 sources, while the remaining three did not show any oxygen in their spectra (see Discussion). The results of our calculation are shown in Table~\ref{tab:oiii}. In the following, we also considered some additional physical properties of the sources such as their bolometric luminosity, radio luminosity, and radio loudness parameter. The bolometric luminosity was estimated using its scaling relation with the H$\beta$ luminosity, as in \citep{Berton15a}. The radio luminosities are derived from B18 and \citep{Berton20b}. The radio loudness parameter is defined as the ratio between the radio flux density at 5~GHz and the flux density in the optical B band. When the ratio is larger than 10, the source is radio loud. Conversely, it is radio quiet \citep{Kellermann89}. To estimate it, we either used the values published from \citep{Foschini15, Berton15a}, or we calculated it using the B-band magnitude values reported in the NASA Extragalactic Database (NED) and our own measurements at 5~GHz. \par

\subsection{Correlations with the whole sample}
\label{sec:whatever}

\begin{table}[t]
\caption{Results of the correlations between the [O~III] core shift and the physical properties of the sources. Columns: (1) First tested sample; (2) second tested sample; (3) Pearson correlation coefficient; (4) p-value of the Pearson coefficient; (5) Spearman's rank correlation coefficient; (6) p-value of the Spearman coefficient.}
\label{tab:correlations}
\centering
\begin{tabular}{lccccc}
\hline
Q1 & Q2 & r$_P$ & p-value(P) & r$_S$ & p-value(S) \\
\hline
L$_{\rm [O~III]}$ & v$_c$ & -0.05 & 0.68 & 0.10 & 0.38 \\
L$_{\rm [O~III]}$ & |v$_c$| & 0.07 & 0.54 & 0.24 & 0.03 \\
L$_{\rm bol}$ & v$_c$ & -0.10 & 0.37 & 0.01 & 0.97 \\
L$_{\rm bol}$ & |v$_c$| & 0.15 & 0.18 & 0.28 & 0.01 \\
L$_{\rm 5~GHz}$ & v$_c$ & 0.01 & 0.97 & 0.01 & 0.92 \\
L$_{\rm 5~GHz}$ & |v$_c$| & 0.03 & 0.78 & 0.39 & 0.0004 \\
RL & v$_c$ & 0.18 & 0.12 & 0.03 & 0.81 \\
RL & |v$_c$| & 0.12 & 0.29 & 0.36 & 0.001 \\
\hline
    \end{tabular}
    \label{tab:corr_lum}
\end{table}

\begin{figure*}[!t]
\centering
    \begin{minipage}[t]{.49\textwidth}
	\centering
	\includegraphics[trim={0cm 0cm 0cm 0cm}, width=\textwidth]{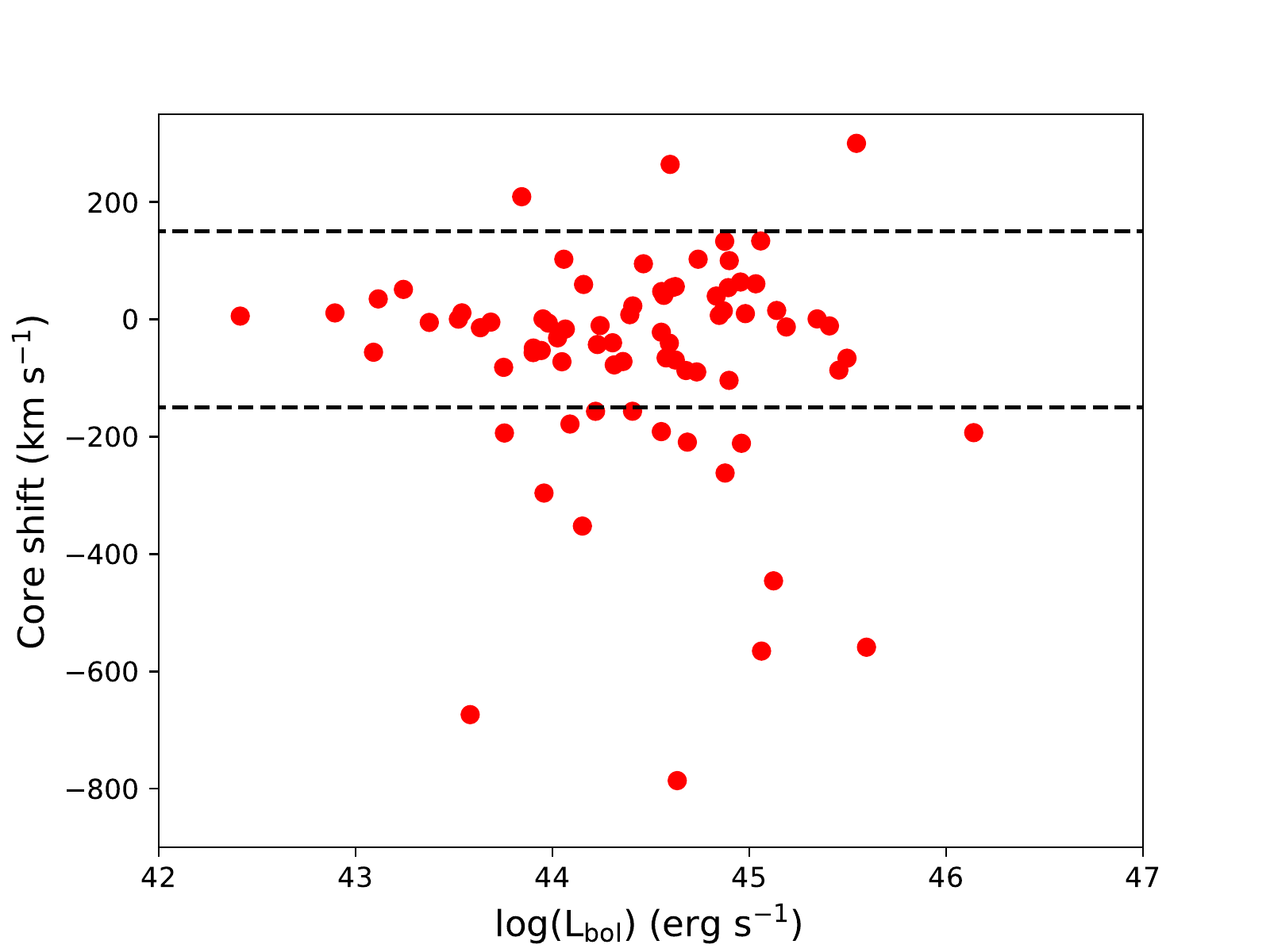}
    \end{minipage}
\hfill
    \begin{minipage}[t]{.49\textwidth}
        \centering
	\includegraphics[trim={0cm 0cm 0cm 0cm}, width=\textwidth]{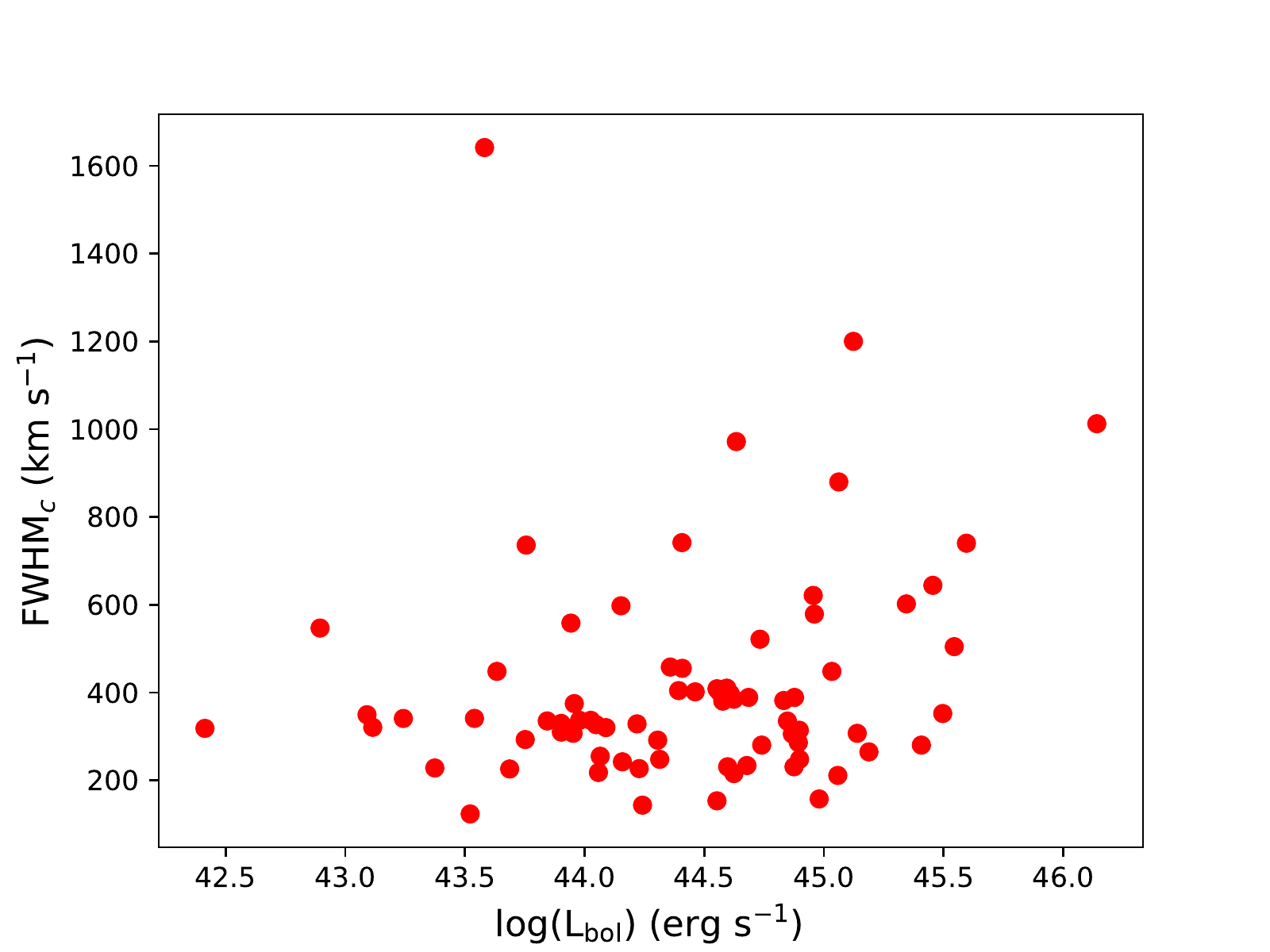}
	\end{minipage}
\caption{\textbf{Left:} [O~III] core velocity shift in \kms\ as a function of the bolometric luminosity in \ergs. The horizontal dashed lines indicate the limits for red or blue outliers at 150 \kms. \textbf{Right:} FWHM([O~III]) core component in \kms\ as a function of the bolometric luminosity in \ergs.}
\label{fig:bol-o3}
\end{figure*}

The property we are most interested in is the shift of the whole [O~III] line, that is the blue outlier phenomenon. Our aim is to have a better understanding of what may be causing this shift. We tested the core velocity against four physical properties that may be associated with it, that is the actual [O~III] line luminosity, the bolometric luminosity, the radio luminosity, and the radio loudness parameter. The first two, indeed, have been found to possibly affect the [O~III] core shift especially at high redshift \citep{Marziani16b}. The radio luminosity and the radio loudness parameter, instead, may be related with the presence of relativistic jets (\citealp[with some caveats, see e.g.,][]{Padovani17, Jarvela17}). The results of these tests are reported in Table~\ref{tab:corr_lum}. 
None of the tested quantities is strongly correlated to the velocity of the [O~III] core. This is true also when the absolute value of the shift is considered, neglecting whether the line is blueshifted or redshifted. An example is shown in Fig.~\ref{fig:bol-o3}. Since it is known that [O~III] shift can also be radiatively driven in high luminosity sources \citep{Marziani16b}, and a weak, but significant correlation is identified by the Spearman's rank correlation coefficient between L$_{\rm bol}$ and |v$_c$|, we wanted to investigate whether an L$_{\rm bol}$ threshold exists above which a source has a higher possibility of showing an [O~III] shift or a higher FWHM([O~III]). As seen in the left panel of Fig.~\ref{fig:bol-o3}, however, no clear threshold seems to be present to produce an [O~III] shift. In the FWHM distribution, the highest values of FWHM are found mostly, but not exclusively, among the highest luminosity sources, although the statistic is too poor to draw any meaningful conclusion. \\
It is also worth mentioning that the Spearman's rank correlation coefficient has a rather low p-value when the radio loudness parameter and, even more noteworthy, the radio luminosity are tested against the absolute value of the core shift velocity, suggesting that these two weak correlations may be significant. The radio luminosity, however, contains some dependence on the redshift. Indeed, B18 already identified a correlation between the radio luminosity and the redshift, present also in our sample (r$_P$ = 0.51, p-value=3$\times10^{-6}$), which is due to the sample selection. For this reason, we carried out a partial correlation analysis, using the redshift as a control parameter. In this case, the Spearman's rank correlation coefficient becomes 0.26 (p-value = 0.03), so the correlation seems to disappear. We carried out the same analysis on the other correlations involving the radio luminosity shown in Table~\ref{tab:corr_lum}, but we found no statistically significant result. \\
The radio loudness parameter instead is redshift independent, and it also shows a weak but statistically significant correlation with the [O~III] core shift. As we already mentioned, radio loudness is often used as a proxy to detect the presence of relativistic jets. This, therefore, suggests that sources harboring relativistic jets may tend to present [O~III] core shift more often than those without a jet. To verify this hypothesis, we decided to split the sample of sources according to the presence or not of a relativistic jet and of other radio properties derived in B18, and to carry out some statistical tests to verify if and how these features affect the optical lines' profile. 

\subsection{Radio classification}
The first radio-based classification, jetted (J) or non-jetted (NJ), is based on the data from B18, where each source was analyzed individually. In general, the classification criteria were the radio luminosity, which is typically higher than 10$^{40}$ \ergs in jetted NLS1s, the spectral index, since a flat radio spectrum with $\alpha_\nu < 0.5$ indicates the presence of relativistic beamed jets, and the multiwavelength properties of the sources.
In previous studies as B16, each source could be classified only according to the radio-loudness parameter \citep{Kellermann89}, defined as the ratio between the flux density at 5~GHz and in the optical B band. Sources with radio-loudness larger than 10 are classified as radio-loud, conversely as radio-quiet. Objects with no previous detection in radio surveys are often labeled as radio-silent. Usually, radio-loud sources harbor relativistic jets, while the origin of radio emission in radio-quiet objects is strongly debated \citep[e.g.,][]{Panessa19}. However, several sources do not obey this rule, therefore a classification criterion based on physical properties is needed \citep{Padovani17}. Based on the radio observations, B18 established whether relativistic jets are present or not in each source. An example is J1038+4227, which was classified as radio-loud by \citet{Foschini15} (and so, estimated to be jetted), but was later found to not have relativistic jets, since the radio emission originates from prominent star forming activity \citep{Caccianiga15}. \\
The second subdivision is based on the radio morphology of our NLS1s. B18 classified them with a compactness criterion. They used the ratio $\mathcal{R} = S_p/S_i$, that is the ratio between the peak flux density and the integrated flux density of the whole emitting region. Sources with $\mathcal{R} \geq 0.95$ were classified as compact (C), those with $0.75 \leq \mathcal{R} < 0.95$ were labeled as intermediate (I), and finally those with $\mathcal{R} < 0.75$ were classified as extended (E). While jetted sources were mostly (but no exclusively) compact, non-jetted sources typically had an extended morphology. The morphological parameters were calculated using the flux densities reported in B18 and \citep{Berton20b}. For the sources without a detection in \citep{Berton20b}, we assumed a compact morphology. It is worth noting that since the radio observations were carried out in the most extended configuration of JVLA, some large-scale extended emission may had been resolved out. However, a sample of NLS1s were investigated at the same frequency by \citet{Chen20} using the JVLA in a more compact configuration. Their results seem to indicate that the morphology of jetted sources is typically compact, with no signs of large-scale radio lobes as those seen in other classes of jetted AGN. Furthermore, the spatial scale of these structures even in A configuration would be several kpc and possibly outside of the host galaxy.\\
The last classification was used only for jetted sources, and is based on the division to sources with flat (F) or steep (S) radio spectral index $\alpha_\nu$ ($F_\nu \propto \nu^{-\alpha_\nu}$). As in other AGN, in jetted NLS1s the radio spectral index is a rather good indicator of the source inclination, that is the angle between the line of sight and the relativistic jet axis. In most AGN the spectrum of the radio core appears flat due to the $\tau$=1 surface, where $\tau$ is the optical depth, moving closer to the BH as a function of frequency \citep{Falcke95}. When our line of sight is close to the axis of the system, the emission of the radio core experiences relativistic beaming, that can result in the radio core emission to dominate the whole spectrum. If we are seeing the radio core at larger angles, the relativistic effects diminish and the contribution of extended, often optically thin, emission becomes more important. A properly formed, powerful jet is often characterized by several components, launched at different times, that each have a convex spectrum, that is synchrotron self-absorbed below the turnover frequency, $\nu_\textrm{t}$, and shows an inverted spectral index ($\alpha_\nu \sim$ -2.5), and optically thin with a spectral index of $\alpha_\nu \sim$ 0.7 above $\nu_\textrm{t}$. As a component ages it loses energy and expands, and the peak moves to lower frequencies. A superposition of several components with different ages results in a seemingly flat, but variable, radio spectrum \citep{Blandford79}. This is especially pronounced in sources whose jet axis is close to our line of sight, since the emission is enhanced by relativistic beaming, and in these sources the beamed jet and radio core emission dominates the whole radio spectrum. In sources that we observe at larger angles, the beaming effects become less significant, and whereas the radio core still appears flat, the overall radio spectrum becomes dominated by extended, optically thin, emission, with a steeper spectral index. The spectral classification used in this work is derived from B18 and J\"arvel\"a et al. (in prep.). All the classes are reported in Table~\ref{tab:sample}.\\

\subsection{Statistical tests}

We studied the properties of the [O~III] lines by grouping them according to the radio properties as described in the previous section. For each distribution, we estimated its median value, its standard deviation, and its interquartile range (IQR). We remark that the standard deviation is not used here as an estimate of the error associated to the median, but instead as a proxy of how much each distribution is concentrated around its median value. Furthermore, to examine whether two distributions are statistically different or not, we used both the Kolmogorov-Smirnov (K-S) and Anderson-Darling (A-D) test. The A-D test is more powerful than the K-S in most cases \citep{Hou09}, but we decided to adopt both tests to provide results comparable to the literature and B16. The null hypothesis of these tests is that the two distributions originate from the same population of sources. In the following, we will reject it only if p-value$\leq$0.05. Finally, to account for the measurement errors, in addition to the observed samples we created simulated samples of sources. In each distribution, to each measurement we added a Gaussian noise whose amplitude was equal to the measurement error, thus creating new samples with the same size as the observed ones. We repeated this operation 100 times and performed every time both statistical tests, to obtain a median p-value for both K-S and A-D test. The goal of these simulations is to verify the stability of the p-values we measured on the observed samples. Therefore, we also estimated the standard deviation of the distributions of p-values obtained from the simulations. \\

\subsection{Jets and [O~III]}
\subsubsection{Line core component}
\begin{figure*}[!t]
\centering
    \begin{minipage}[t]{.49\textwidth}
	\centering
	\includegraphics[trim={0cm 0cm 0cm 0cm}, width=\textwidth]{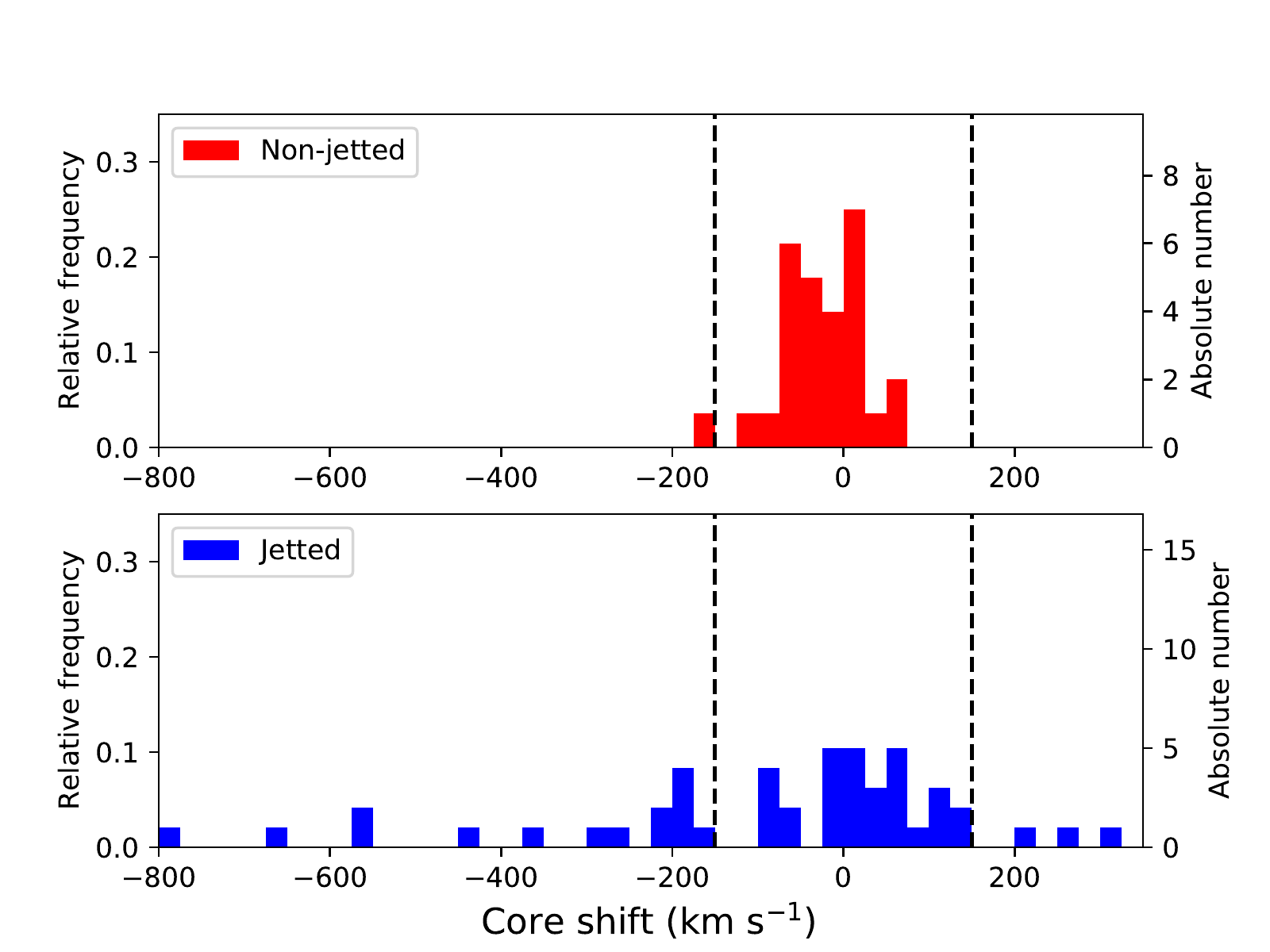}
    \end{minipage}
\hfill
    \begin{minipage}[t]{.49\textwidth}
        \centering
	\includegraphics[trim={0cm 0cm 0cm 0cm}, width=\textwidth]{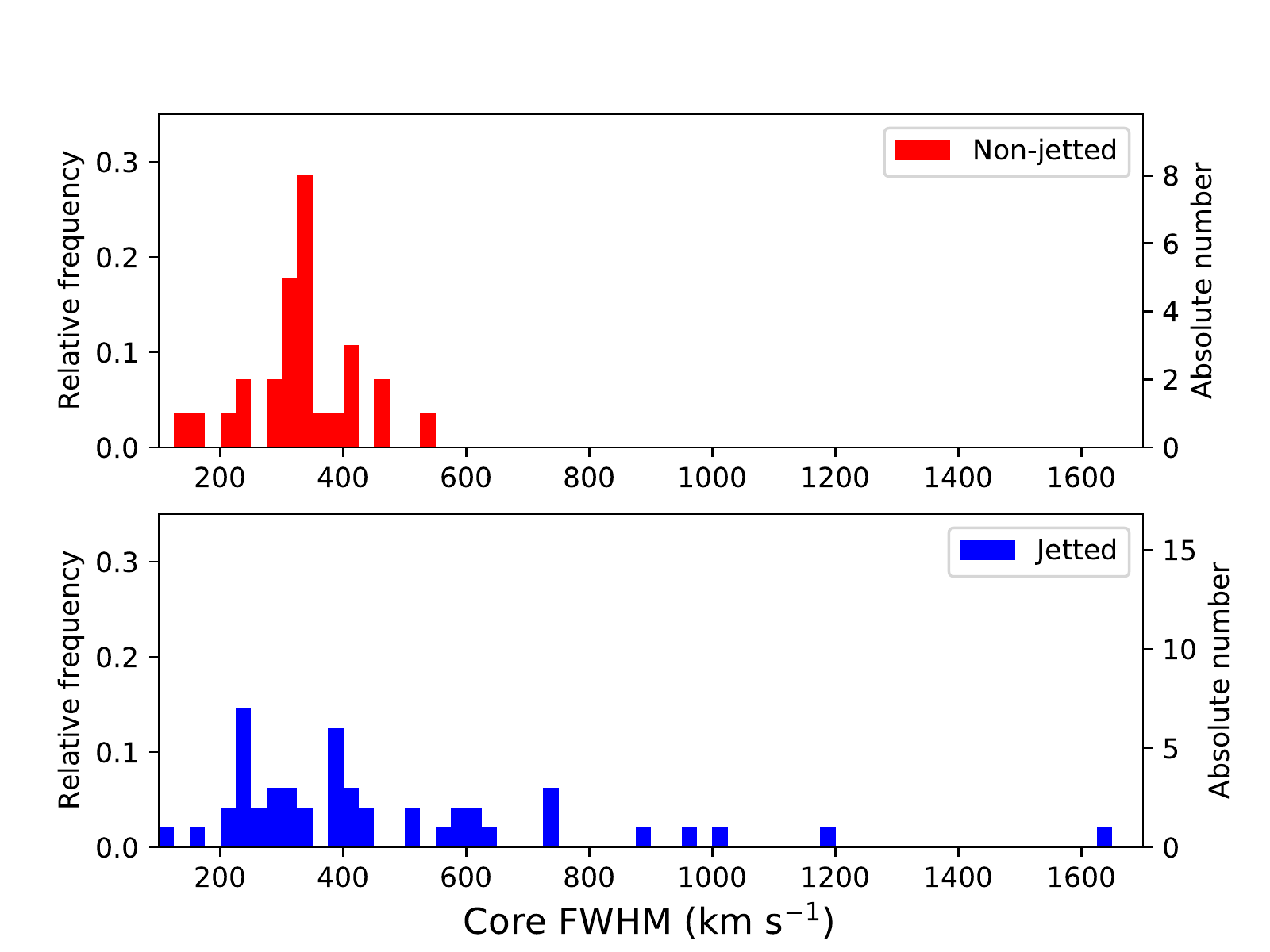}
	\end{minipage}
\caption{\textbf{Left:} distributions of the [O~III] core velocity shift in \kms, in non-jetted (top panel) and jetted sources (bottom panel). The vertical dashed lines indicate the limits for red or blue outliers at 150 \kms. Binning of 25 \kms. \textbf{Right:} distributions of the FWHM([O~III]) core component in \kms, in non-jetted (top panel) and jetted sources (bottom panel). Binning of 25 \kms.}
\label{fig:blue_outliers}
\end{figure*}   

\begin{table}[t]
\caption{Results of the statistical tests for the spectral index samples. Columns: (1) Tested sample; (2) median velocity; (3) standard deviation of the velocity; (4) interquartile range; (5) p-value of the K-S tests against the other sample, observed in top row, simulated in middle row, and standard deviation of the simulated p-value in the bottom row; (6) p-value of the A-D tests against the other sample, observed in top row, simulated in middle row, and standard deviation of the simulated p-value in the bottom row.}
\centering
\begin{tabular}{l|ccc|ccccccccccccc}
\hline
Sample & \makecell{Median \\ (\kms) } & \makecell{Std \\ (\kms) } & \makecell{IQR \\ (\kms) } & \makecell{K-S$^{obs}$ \\ K-S$^{sim}$ \\ K-S$^{std}$} & \makecell{A-D$^{obs}$ \\ A-D$^{sim}$ \\ A-D$^{std}$} \\
\hline
v$_c$(NJ) & -26.5 & 48.5 & 71.0 & \makecell{0.06 \\ 0.06 \\ 0.02} & \makecell{0.01 \\ 0.02 \\ 0.01} \\
\hline
v$_c$(J) & -11.9 & 226.4 & 249.1 & \makecell{0.06 \\ 0.06 \\ 0.02} & \makecell{0.01 \\ 0.02 \\ 0.01} \\
\hline
FWHM$_c$(NJ) & 329.4 & 85.9 & 69.1 & \makecell{0.05 \\ 0.03 \\ 0.02} & \makecell{0.03 \\ 0.02 \\ 0.01} \\
\hline
FWHM$_c$(J) & 385.7 & 289.8 & 328.3 & \makecell{0.05 \\ 0.03\\ 0.02} & \makecell{0.03 \\ 0.02 \\ 0.01} \\
\hline
v$_w$(NJ) & -268.9 & 127.7 & 98.8 & \makecell{0.06 \\ 0.10 \\ 0.09} & \makecell{0.05 \\ 0.07 \\ 0.05} \\
\hline
v$_w$(J) & -335.7 & 244.1 & 298.6 & \makecell{0.06 \\ 0.10 \\ 0.09} & \makecell{0.05 \\ 0.07 \\ 0.05} \\
\hline
FWHM$_w$(NJ) & 868.6 & 346.2 & 311.4 & \makecell{0.03 \\ 0.06 \\ 0.05} & \makecell{0.07 \\ 0.07 \\ 0.04} \\
\hline
FWHM$_w$(J) & 1180.3 & 427.8 & 607.7 & \makecell{0.03 \\ 0.06 \\ 0.05} & \makecell{0.07 \\ 0.07 \\ 0.04} \\
\hline
\end{tabular}
\label{tab:summary_jets}
\end{table} 

\begin{figure}[!t]
\centering
\includegraphics[width=0.6\hsize]{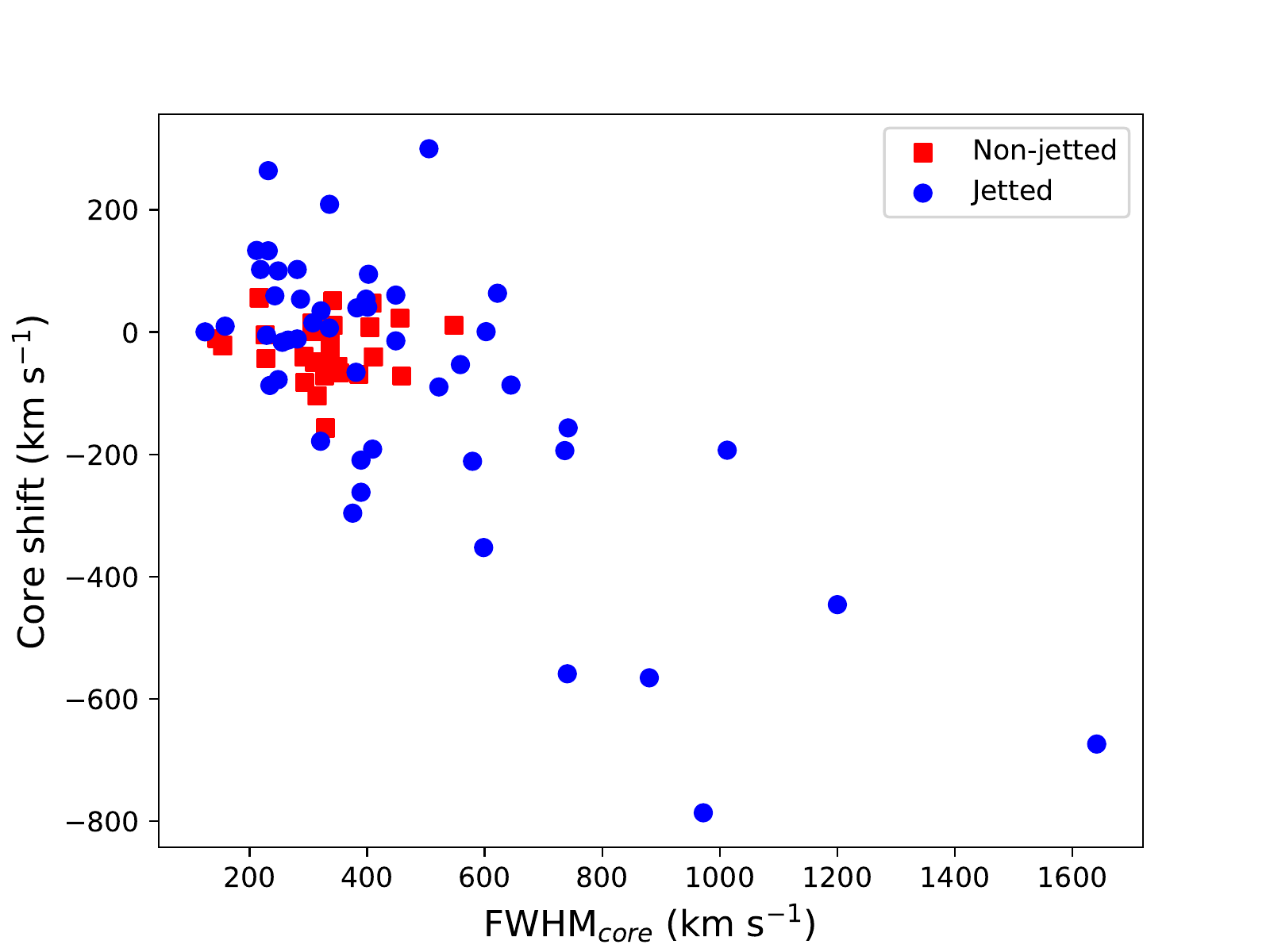}
\caption{Correlation between the FWHM of the [O~III] core component and its shift. Non-jetted sources are represented with the red squares, while jetted sources are indicated with the blue circles.}
\label{fig:core-fwhm}
\end{figure}

First of all, we examined the properties of the [O~III] lines separately in jetted and non-jetted sources, as indicated in the second to last column of Table~\ref{tab:sample}. We first compared the velocity of the core component of [O~III] with respect to the galaxy rest frame in the two samples. We remark that, by definition, a source is considered a red or blue outlier when the velocity of the line core component is $|v_c| >150$ \kms, following the definition by \citet{Komossa08}. The distributions are shown in the left panel of Fig.~\ref{fig:blue_outliers}. Among our 28 non-jetted sources, we found only one blue outlier ($\sim$4\%) and no red outliers. In our sample of 48 jetted sources, instead, we found 15 blue outliers ($\sim$31\%) and 3 red outliers ($\sim$6\%). \\
The statistical properties of the two distributions are shown in Table~\ref{tab:summary_jets}. It is evident that while the median values of the two samples are similar, both the standard deviation and the IQR are very different. Both quantities are much larger in jetted sources, suggesting that the [O~III] core velocity distribution in this subsample is significantly more spread out. Despite this difference, the K-S test indicates that the null hypothesis cannot be rejected, but the A-D test provides the opposite result (p-value = 0.01). We repeated the tests on the simulated samples, and the results did not change significantly. However, since our goal is to understand the nature of blue outliers, and they are concentrated in the distributions' tail, the A-D test is more reliable than the K-S test. We can therefore conclude that the null hypothesis can be rejected, and that the core shift distributions do not originate from the same population of sources. \\
We then tested the distributions of FWHM of the [O~III] core component in the two samples, shown in the right panel of Fig.~\ref{fig:blue_outliers}. As we did before, we tested the differences between the two samples, and also in this case the distribution of jetted sources has larger standard deviation and IQR. In this case, both statistical tests suggest that the null hypothesis can be rejected in the observed samples, and also the simulated ones. Therefore, we can conclude that the two distributions of [O~III] core FWHM do not originate from the same population of sources. \\
We also studied whether a correlation is present among the core shift and the FWHM of the [O~III] core component. The result is shown in Fig.~\ref{fig:core-fwhm}, while the values we obtained are reported in Table~\ref{tab:correlations}. The Pearson correlation coefficient for all sources is $r=-0.70$, and its p-value, that is the probability that this result is due to a random distribution of sources, is 8$\times10^{-13}$. However, as we have seen before, a core shift is present almost exclusively in jetted sources. Therefore, we decided to test the two samples separately. In non-jetted sources, no correlation is present ($r=0.02$, p-value = 0.93), indicating that the FWHM of the [O~III] core and its position are independent of each other. Conversely, in jetted sources the Pearson correlation coefficient is higher ($r = -0.73$), and the p-value is $4\times10^{-9}$, indicating that the correlation is present only in this sample of objects. It is worth noting that this correlation is utterly dominated by the presence of the blue outliers. Indeed, when the Pearson coefficient is evaluated in this sample rejecting the blue outliers, its value becomes $r=-0.28$ (p-value = 0.13). Vice versa, when only the blue outliers are tested, the correlation remains, although with a lower p-value (r=$-0.67$, p-value = $3\times10^{-3}$). In conclusion, [O~III] core velocity and FWHM do not seem to correlate in the bulk of NLS1s population, but they are related among blue outliers. \\

\subsubsection{Wing component}
\begin{figure*}[!t]
\centering
    \begin{minipage}[t]{.49\textwidth}
	\centering
	\includegraphics[trim={0cm 0cm 0cm 0cm}, width=\textwidth]{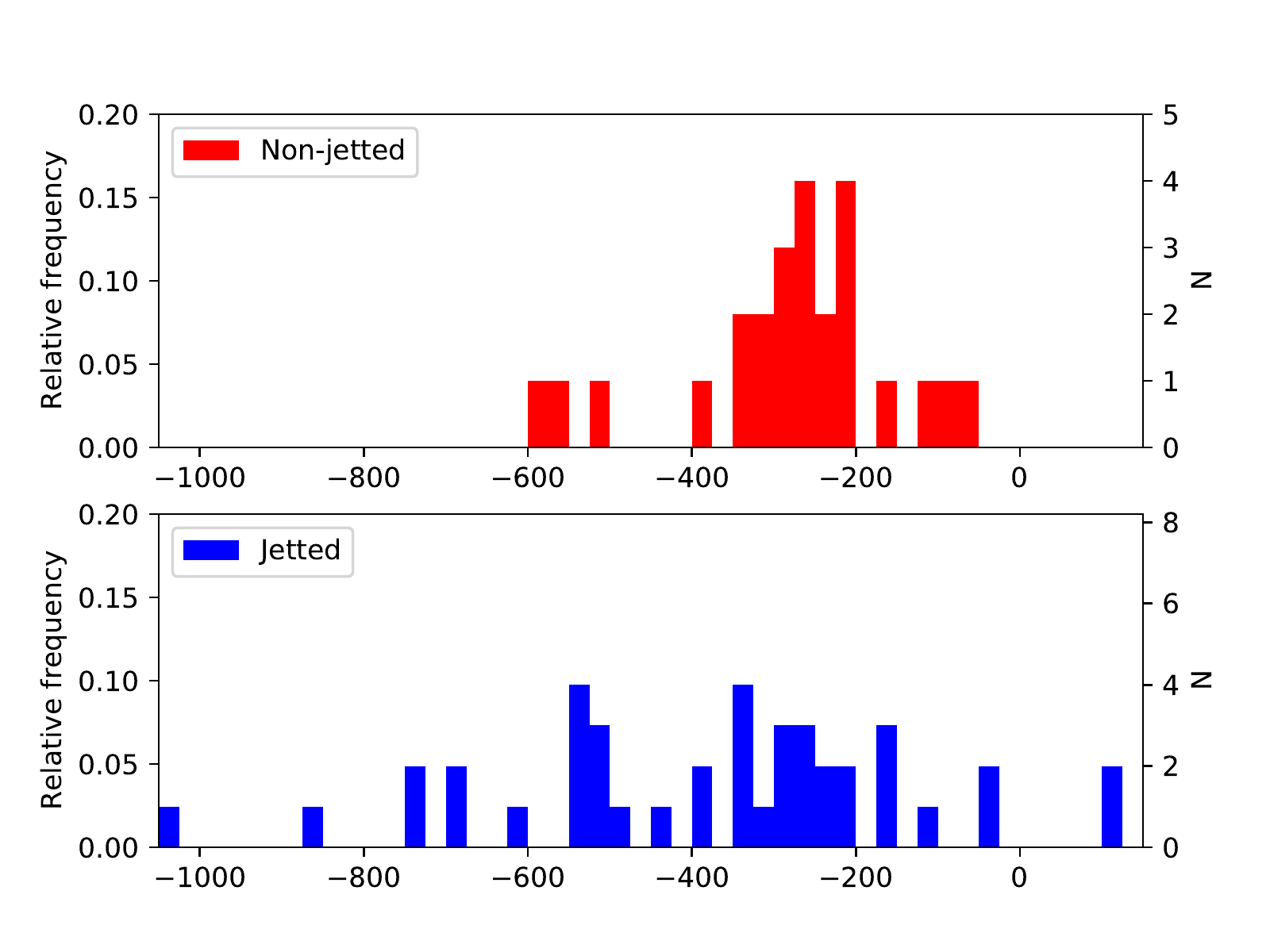}
    \end{minipage}
\hfill
    \begin{minipage}[t]{.49\textwidth}
        \centering
	\includegraphics[trim={0cm 0cm 0cm 0cm}, width=\textwidth]{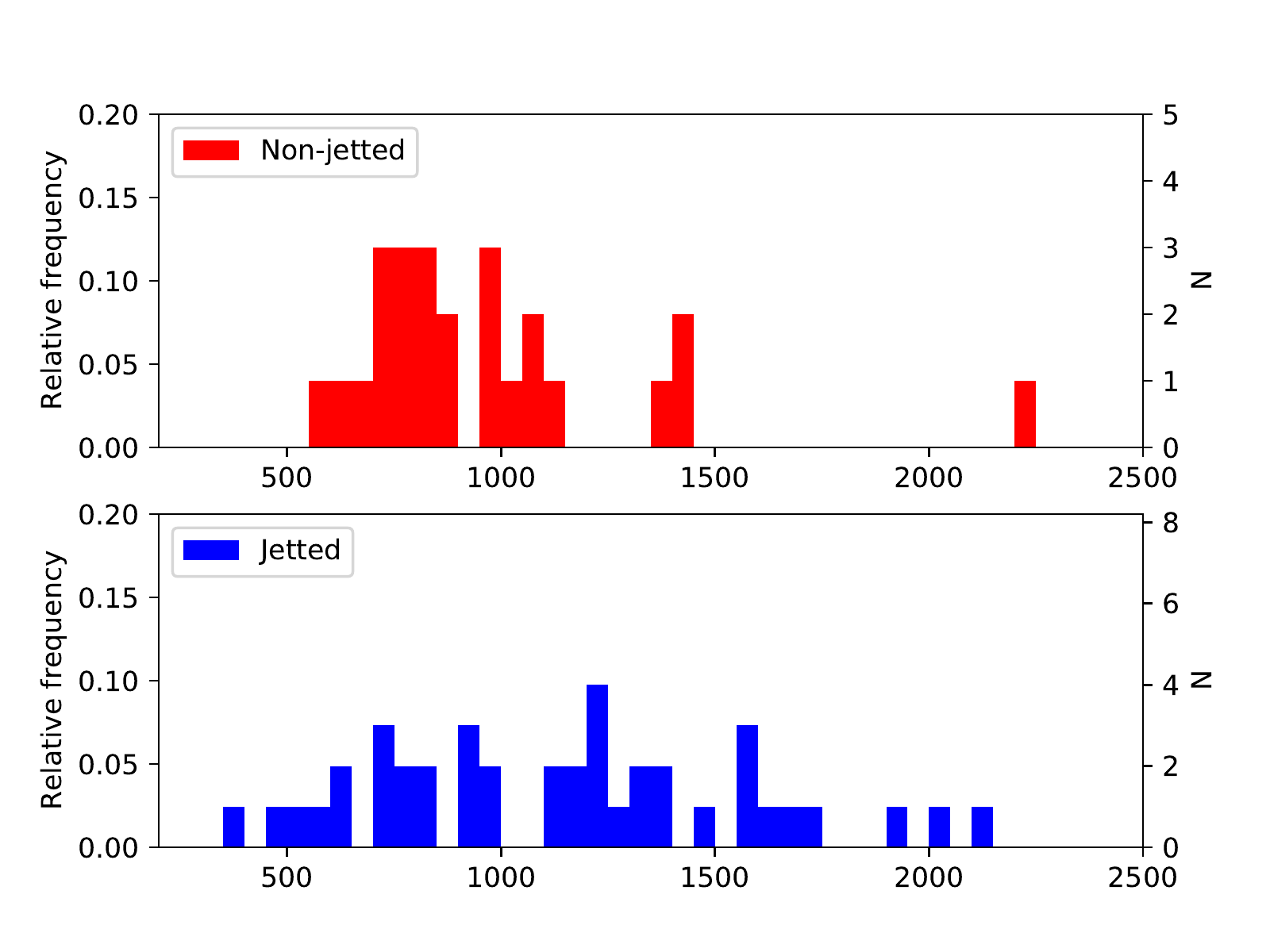}
	\end{minipage}
\caption{\textbf{Left:} distributions of the [O~III] wing velocity shift in \kms, in non-jetted (top panel) and jetted sources (bottom panel). Binning of 25 \kms. \textbf{Right:} distributions of the FWHM([O~III]) wing component in \kms, in non-jetted (top panel) and jetted sources (bottom panel). Binning of 50 \kms.}
\label{fig:jet-wings}
\end{figure*}

As previously mentioned, the wing component is likely associated to an outflow coming from the inner part of the NLR. A wing component could be identified during the decomposition process in all but ten sources, three non-jetted and seven jetted. We tested the distribution of wing velocity in jetted and non-jetted sources, shown in the left panel of Fig.~\ref{fig:jet-wings}, and whose parameters are summarized in Table~\ref{tab:summary_jets}. As for the distributions of the [O~III] core properties, the wing velocity distribution is somewhat peaked in non-jetted sources, with a large majority of sources having velocities between -400 and -200 \kms, while it is more spread out in the jetted sample. 

However, the statistical tests seem to indicate that the null hypothesis cannot be rejected with the required confidence level. The same is true for the distributions of wing FWHM, displayed in the right panel of Fig.~\ref{fig:jet-wings}. The presence of relativistic jets does not ultimately seem to affect the FWHM of the [O~III] wing, which remains as turbulent in the jetted NLS1s as in the non-jetted NLS1s. However, it is worth noting that the standard deviation of the p-value derived in the simulations is rather large because of the relatively large measurement errors on the wing properties. Therefore, our data may not be enough to reach a robust conclusion. \\
Finally, to better understand if and how the physical quantities we measured are related to each other, we tested various correlations by evaluating the Pearson coefficients. The results are shown in Table~\ref{tab:correlations}. We first tested the presence of a correlation between the wing velocity and both [O~III] core velocity and FWHM velocity, but we do not find any strong correlation, except for a trend between the wing velocity and the [O~III] core FWHM in jetted sources (r = -0.50, p-value = 8$\times10^{-4}$). In a similar fashion, the FWHM of the wing component does not appear to correlate with any property of the [O~III] core component, except again for the [O~III] core FWHM in jetted sources (r = 0.54, p-value = 2$\times10^{-4}$). We finally tested the presence of a correlation between the wing velocity and its FWHM, and we find another possible trend among these two quantities in non-jetted sources (r = -0.51, p-value = 0.01). 
 
\subsection{Radio morphology and [O~III]}
\begin{figure*}[!t]
\centering
    \begin{minipage}[t]{.49\textwidth}
	\centering
	\includegraphics[trim={0cm 0cm 0cm 0cm}, width=\textwidth]{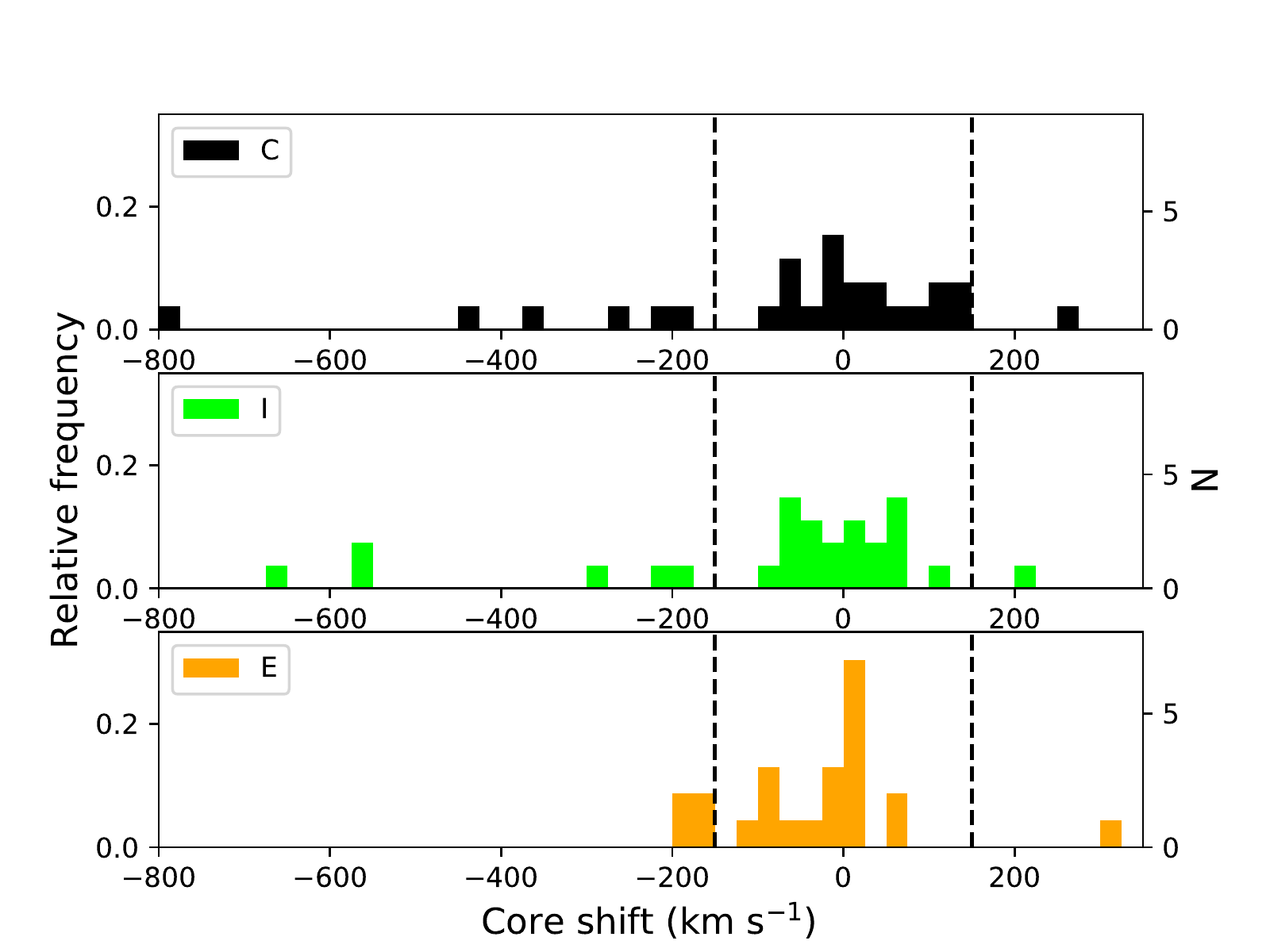}
    \end{minipage}
\hfill
    \begin{minipage}[t]{.49\textwidth}
        \centering
	\includegraphics[trim={0cm 0cm 0cm 0cm}, width=\textwidth]{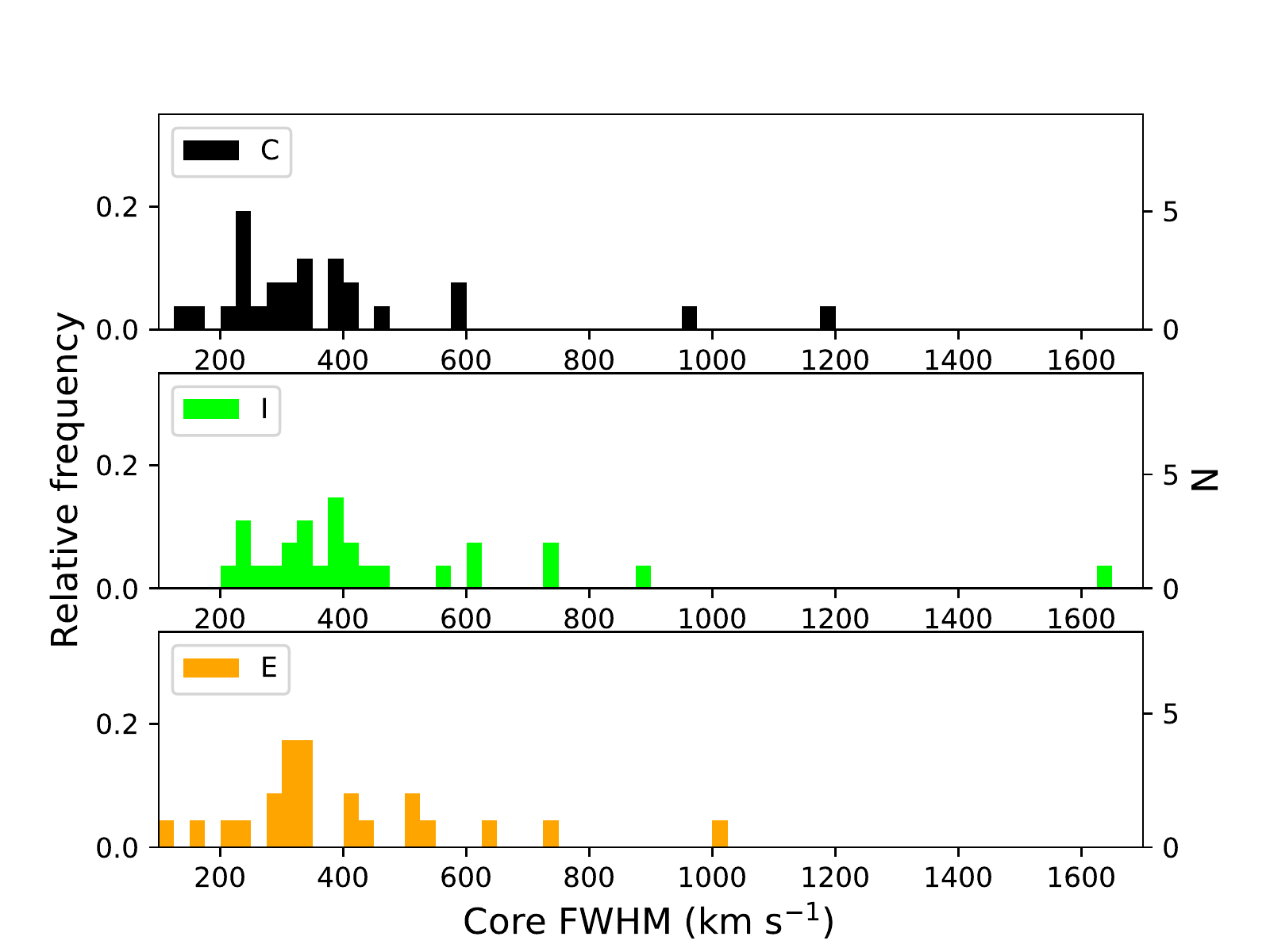}
	\end{minipage}
\caption{\textbf{Left:} distributions of the [O~III] core velocity shift in \kms, in compact (top panel), intermediate (middle panel) and extended sources (bottom panel). The vertical dashed line indicate the limit for red or blue outliers at 150 \kms. Binning of 25 \kms. \textbf{Right:} distributions of the FWHM([O~III]) core component in \kms\ in the same samples. Binning of 25 \kms.}
\label{fig:morph-core}
\end{figure*}

We tried to investigate whether the radio morphology is somehow connected to the [O~III] properties. 
As before, we first analyzed the [O~III] core velocity distributions, which are shown in the left side of Fig.~\ref{fig:morph-core}. Outliers do not seem to appear preferentially in any specific sample. Among the 26 C sources, we found one red outlier and six blue, among 27 I sources there is one red outlier and six blue, and finally in E sources we found one red outlier and four blue. To check whether the distributions were different from each other, we used the K-S and A-D tests both on the observed samples and on the simulated samples. All tests, even when accounting for the p-values distributions, confirm that the null hypothesis can never be rejected. The results of all tests, as well as the median velocity, its standard deviation, and the IQR are all reported in Table~\ref{tab:summary_morph}. 

\begin{table}[!h]
\caption{Statistical tests results for the morphological samples. Columns: (1) Tested sample; (2) median velocity; (3) standard deviation of the velocity; (4) interquartile range; (5) p-value of the K-S tests against the corresponding C sample, observed in top row, simulated in middle row, and standard deviation of the simulated p-value in bottom row; (6) p-value of the A-D tests against the corresponding C sample, observed in top row, simulated in middle row, and standard deviation of the simulated p-value in bottom row; (7) K-S tests against the corresponding I sample, rows as before; (8) A-D tests against the corresponding I sample, rows as before; (9) K-S tests against the corresponding E sample, rows as before; (10) A-D tests against the corresponding E sample, rows as before.}
\centering
\footnotesize
\begin{tabular}{l|ccc|cc|cc|ccccccccc}
\hline
Sample & \makecell{Median \\ (\kms) } & \makecell{Std \\ (\kms) } & \makecell{IQR \\ (\kms) } & \makecell{K-S$^{obs}_C$ \\ K-S$^{sim}_C$ \\ K-S$^{std}_C$} & \makecell{A-D$^{obs}_C$ \\ A-D$^{sim}_C$ \\ A-D$^{std}_C$} & \makecell{K-S$^{obs}_I$ \\ K-S$^{sim}_I$ \\ K-S$^{std}_I$} & \makecell{A-D$^{obs}_I$ \\ A-D$^{sim}_I$ \\ A-D$^{std}_I$} & \makecell{K-S$^{obs}_E$ \\ K-S$^{sim}_E$ \\ K-S$^{std}_E$} & \makecell{A-D$^{obs}_E$ \\ A-D$^{sim}_E$ \\ A-D$^{std}_E$} \\
\hline
v$_c$(C) & -11.6 & 210.4 & 125.7 & $-$ & $-$ & \makecell{0.87 \\ 0.87 \\ 0.12} & \makecell{0.89 \\ 0.74 \\ 0.11 } & \makecell{0.36 \\ 0.33 \\ 0.15} & \makecell{0.29 \\ 0.30 \\ 0.08} \\
\hline
v$_c$(I) & -31.3 & 207.2 & 116.1 & \makecell{0.87 \\ 0.87 \\ 0.12} & \makecell{0.89 \\ 0.74  \\ 0.11 } & $-$ & $-$ & \makecell{0.42 \\ 0.52 \\ 0.18} & \makecell{0.32 \\ 0.36 \\ 0.10} \\
\hline
v$_c$(E) & -11.0 & 101.0 & 97.0 & \makecell{0.36 \\ 0.33 \\ 0.15} & \makecell{0.29 \\ 0.30 \\ 0.08} & \makecell{0.42 \\ 0.52 \\ 0.18} & \makecell{0.32 \\ 0.36 \\ 0.10} & $-$ & $-$ \\
\hline
FWHM$_c$(C) & 323.8 & 232.2 & 147.3 & $-$ & $-$ & \makecell{0.36 \\ 0.20 \\ 0.18} & \makecell{0.15 \\ 0.15 \\ 0.09} & \makecell{0.63 \\ 0.60 \\ 0.15} & \makecell{0.65 \\ 0.61 \\ 0.12} \\
\hline
FWHM$_c$(I) & 385.0 & 285.0 & 193.4 & \makecell{0.36 \\ 0.20 \\ 0.18} & \makecell{0.15 \\ 0.15 \\ 0.09} & $-$ & $-$ & \makecell{0.64 \\ 0.43 \\ 0.20} & \makecell{0.56 \\ 0.57 \\ 0.12} \\
\hline
FWHM$_c$(E) & 341.2 & 194.6 & 176.3 & \makecell{0.63 \\ 0.60 \\ 0.15} & \makecell{0.65 \\ 0.61 \\ 0.12} & \makecell{0.64 \\ 0.43 \\ 0.20} & \makecell{0.56 \\ 0.57 \\ 0.12} & $-$ & $-$ \\
\hline
v$_w$(C) & -309.8 & 191.9 & 272.2 & $-$ & $-$ & \makecell{0.95 \\ 0.69 \\ 0.23} & \makecell{0.92 \\ 0.70 \\ 0.21} & \makecell{0.68 \\ 0.44 \\ 0.28} & \makecell{0.54 \\ 0.57 \\ 0.25} \\
\hline
v$_w$(I) & -302.4 & 229.5 & 276.6 & \makecell{0.95 \\ 0.69 \\ 0.23} & \makecell{0.92 \\ 0.70 \\ 0.21} & $-$ & $-$ & \makecell{0.59 \\ 0.59 \\ 0.24} & \makecell{0.60 \\ 0.55 \\ 0.24} \\
\hline
v$_w$(E) & -270.3 & 210.8 & 199.1 & \makecell{0.68 \\ 0.44 \\ 0.28} & \makecell{0.54 \\ 0.57 \\ 0.25} & \makecell{0.59 \\ 0.59 \\ 0.24} & \makecell{0.60 \\ 0.55 \\ 0.24} & $-$ & $-$ \\
\hline
FWHM$_w$(C) & 965.2 & 425.0 & 472.4 & $-$ & $-$ & \makecell{0.65 \\ 0.89 \\ 0.14} & \makecell{0.81 \\ 0.93 \\ 0.17} & \makecell{1.00 \\ 0.99 \\ 0.05} & \makecell{1.00 \\ 1.00 \\ 0.10} \\
\hline
FWHM$_w$(I) & 1082.1 & 404.8 & 561.9 & \makecell{0.65 \\ 0.89 \\ 0.14} & \makecell{0.81 \\ 0.93 \\ 0.17} & $-$ & $-$ & \makecell{0.75 \\ 0.93 \\ 0.16} & \makecell{0.89 \\ 0.84 \\ 0.15} \\
\hline
FWHM$_w$(E) & 999.8 & 383.0 & 567.4 & \makecell{1.00 \\ 0.99 \\ 0.05} & \makecell{1.00 \\ 1.00 \\ 0.10} & \makecell{0.75 \\ 0.93 \\ 0.16} & \makecell{0.89 \\ 0.84 \\ 0.15} & $-$ & $-$ \\
\hline
\end{tabular}
\label{tab:summary_morph}
\end{table}\noindent

We also tested the distributions of FWHM of the [O~III] core component across the different morphological samples. The results are shown in the right side of Fig.~\ref{fig:morph-core}. As for the [O~III] core velocity, however, this parameter does not seem to be related to the radio morphology. All the tests we carried out, and that are summarized in Table~\ref{tab:summary_morph}, seem to point out that the distributions of FWHM are essentially the same across the three samples. Indeed, median values, standard deviation and IQR are all very similar. \\
There is only one minor detail possibly worth noting, that is the blue outliers in the E sample are those with velocity closest to the threshold. Indeed, the [O~III] core velocity distribution of E sources has the lowest standard deviation and IQR. Since most sources with extended morphology do not have relativistic jets, this result could be a (weak) reflection of the difference between the distributions of jetted and non-jetted sources we found in the previous section. However, the samples we are examining do not allow us to draw any statistically robust conclusions. \\
The behavior of [O~III] wings, as for the core component, does not seem to be related in any way to the radio morphology. We tested the wing velocity and FWHM among the three samples, and found that there is no evidence against the null hypothesis. Detailed results are shown in Table~\ref{tab:summary_morph}. 

\subsection{Effect of the redshift}
\begin{figure*}[!ht] 
\centering
    \begin{minipage}[t]{.49\textwidth}
	\centering
	\includegraphics[trim={0cm 0cm 0cm 0cm}, width=\textwidth]{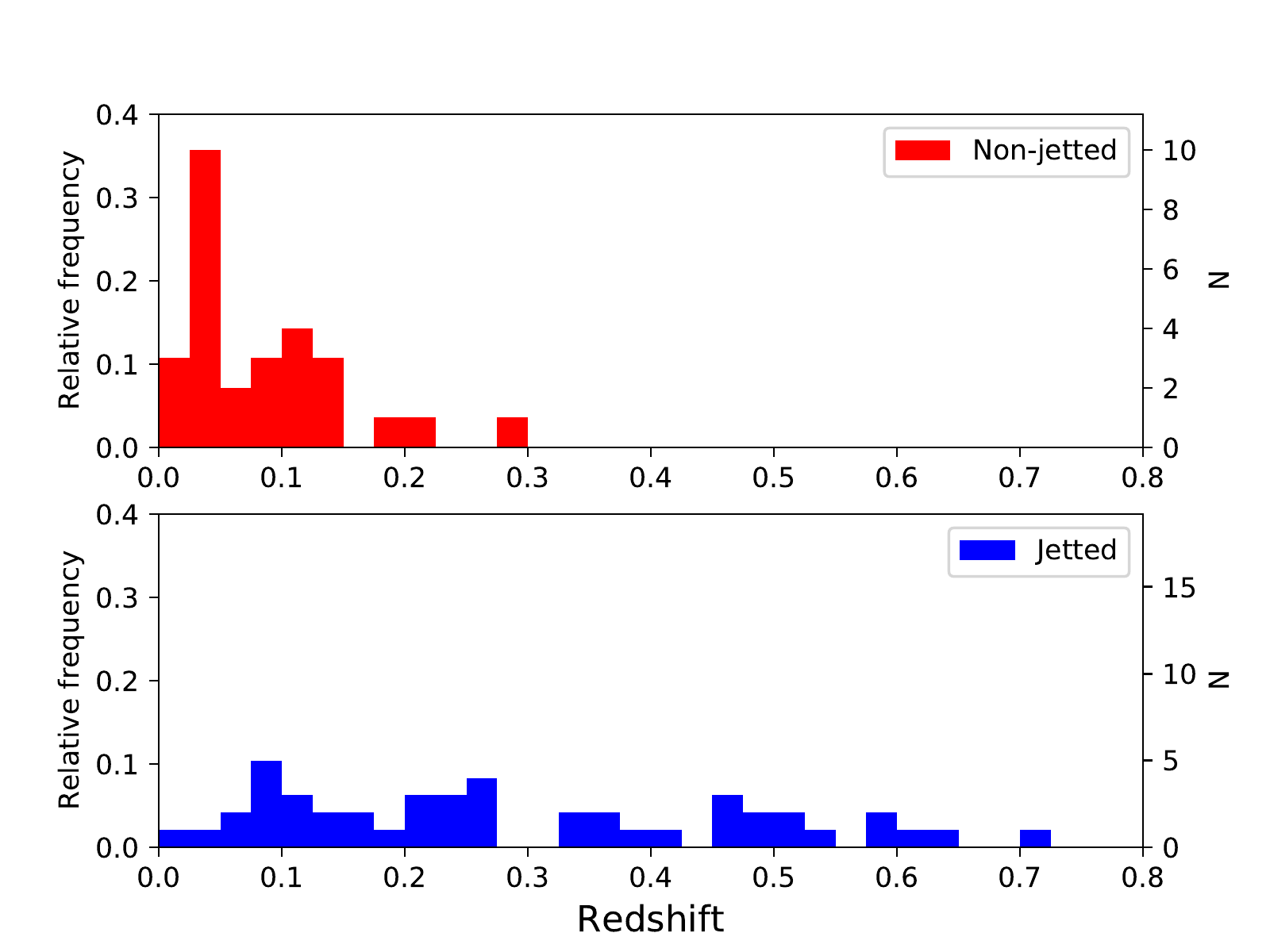}
    \end{minipage}
\hfill
    \begin{minipage}[t]{.49\textwidth}
        \centering
	\includegraphics[trim={0cm 0cm 0cm 0cm}, width=\textwidth]{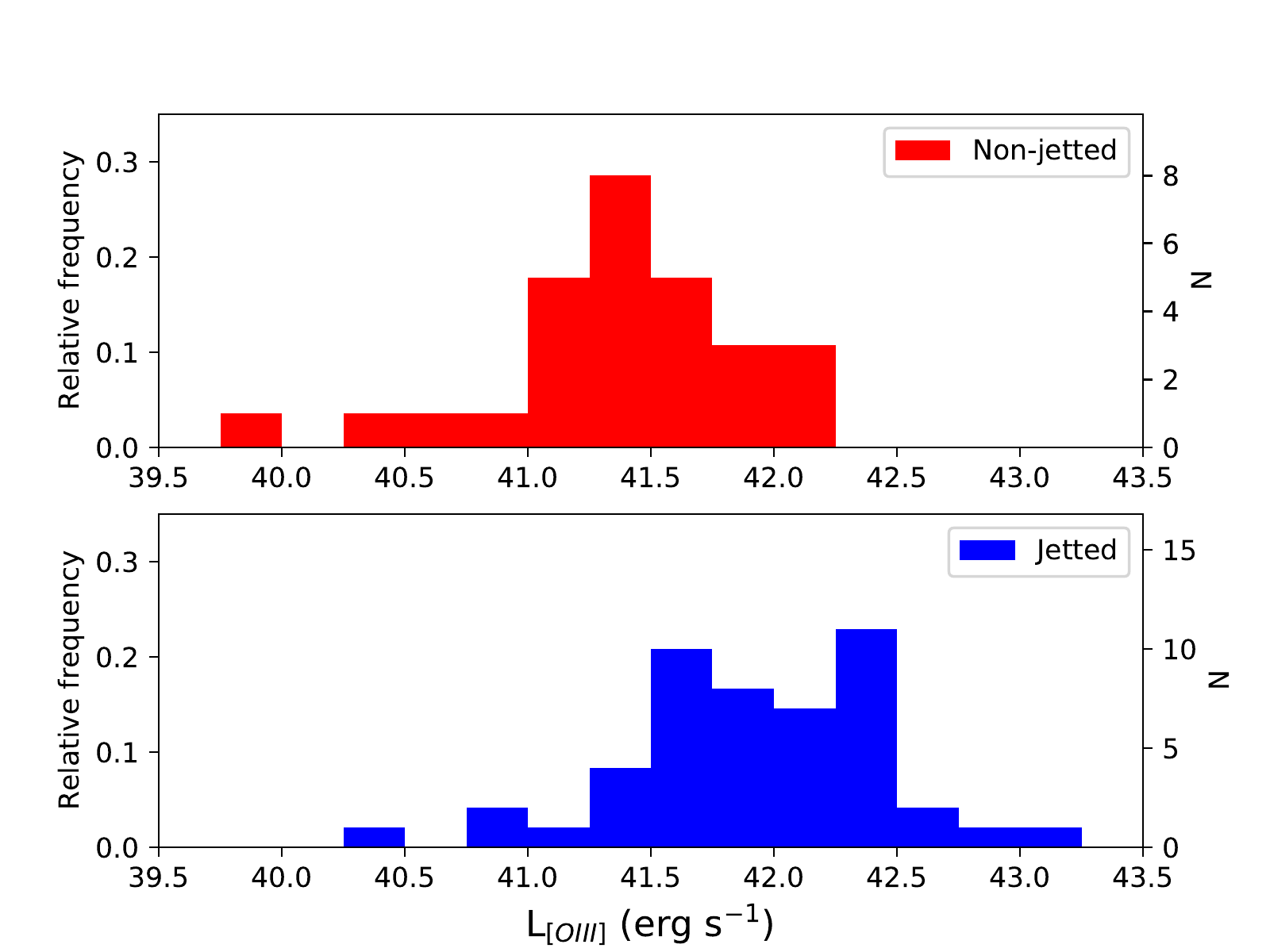}
	\end{minipage}
\caption{\textbf{Left:} \textbf{Left:} Redshift distribution of the jetted (top panel) and the non-jetted (bottom panel) sources. Binning of 0.025. \textbf{Right:} distributions of the logarithm of the [O~III] luminosity (in \ergs) for the jetted (top panel) and the non-jetted (bottom panel) sources. Binning of 0.25 dex.}
\label{fig:red-lum}
\end{figure*}\noindent

\begin{figure}[!]
\centering
\includegraphics[width=0.6\hsize]{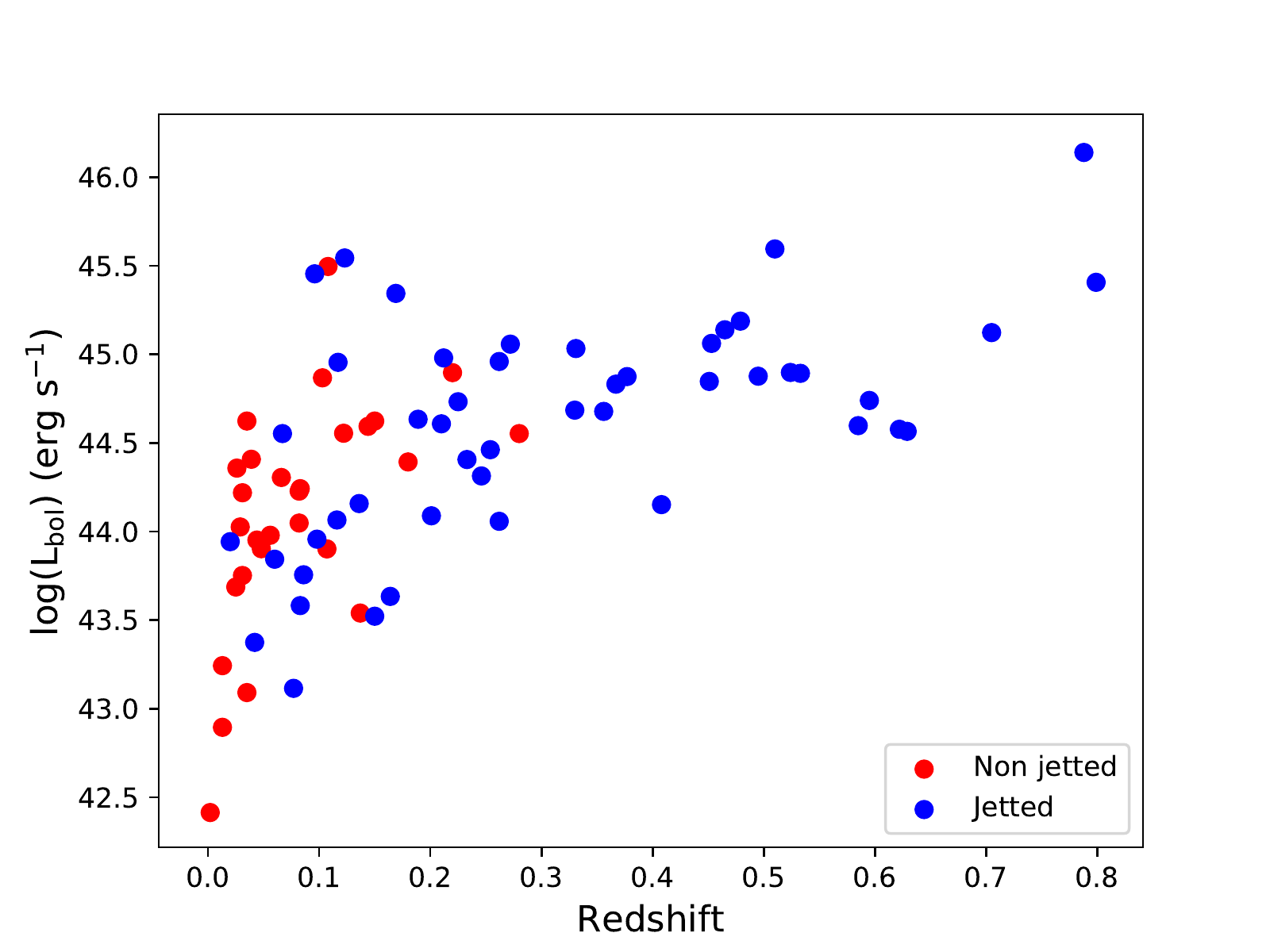}
\caption{The bolometric luminosity of our sources as a function of their redshift. Non-jetted objects are shown in red, while jetted sources are plotted in blue.}
\label{fig:bolz}
\end{figure}

Before moving any further, we tried to quantify the impact that the redshift could have on some possibly redshift-dependent properties. The redshift distribution of jetted and non-jetted sources is indeed rather different, as shown in the left panel of Fig.~\ref{fig:red-lum}, with jetted sources detected up to higher redshift. As shown in B18, the same effect is present in the morphological samples, since sources with extended morphologies are usually not detected at high redshift due to their low surface brightness densities. 
We do not expect significant cosmological evolution of the sources within the redshift range of our samples, both jetted and non-jetted. Indeed, in the radio luminosity function of jetted NLS1s, studied by \citep{Berton16c}, no evolution is observed up to $z$ = 0.6. Only five out of 48 of our jetted sources lie above this threshold, therefore we do not expect them to affect significantly our results. Furthermore, even if some galaxy evolution were present, it is unlikely to affect the interaction between the relativistic jets and the NLR, since the physical mechanisms at play remain the same.\par
We tried also to verify whether the distribution of [O~III] luminosity is different among the jetted and non-jetted sources. Indeed, since jetted sources are located at higher redshift, there could be a selection bias toward more luminous sources. The two distributions are shown in the right panel of Fig.~\ref{fig:red-lum}, and they are actually different, as both K-S and A-D test show p-values$\sim10^{-4}$. However, we tried to verify whether the [O~III] luminosity within the jetted sample is dependent on the redshift, and we found that only a weak correlation is present (r = 0.39, p-value = 0.01). It is also worth noting that, at z$<$0.6, even this weak correlation disappears. Sources with bright [O~III], therefore, are not necessarily located at higher redshift. This suggests that the different luminosity observed in jetted and non-jetted sources is not a selection effect, but a real physical difference between the two classes. This should not come as a surprise, since the presence of a correlation between the [O~III] luminosity and the radio luminosity, which in turn is related to the presence of a relativistic jet, has been known since the seventies \citep{deBruyn78}. \\
We investigate whether it is possible that the higher [O~III] luminosity instead of the relativistic jets affects the presence of blue outliers. A higher line luminosity could originate because of a higher disk luminosity which, in turn, may induce outflows via radiation pressure, and disturbed and asymmetric profiles are often observed in the [O~III] and C IV lines of high-luminosity sources \citep{Marziani16, Marziani16b, Sulentic17}. However, this may not be the case for our sources. As found already by B16, and shown by the correlation coefficients of Tables~\ref{tab:corr_lum} and \ref{tab:correlations}, there is no evidence that the [O~III] core shift correlates with the [O~III] line luminosity, neither when the whole sample nor when the jetted and non jetted subsamples are considered. We also investigated the redshift dependency of the bolometric luminosity. Naturally, a strong correlation is present (see Fig.~\ref{fig:bolz}) with higher luminosity sources seen at higher redshift (Spearman's r$_s$ = 0.67, p-value 3 $\times$ 10$^{-11}$). However, as seen in Sec.~\ref{sec:whatever}, the bolometric luminosity also does not appear to correlate with the [O~III] line properties. Therefore, the presence of the relativistic jets remains the most likely explanation for blue outliers and the other observed properties, at least for the sources included in our sample.

\subsection{[O~III] vs spectral index}
Testing the [O~III] properties as a function of the spectral index will allow us to understand, at least at a first approximation, whether they are dependent on the source inclination or not. 
\begin{figure*}[!ht] 
\centering
    \begin{minipage}[t]{.49\textwidth}
	\centering
	\includegraphics[trim={0cm 0cm 0cm 0cm}, width=\textwidth]{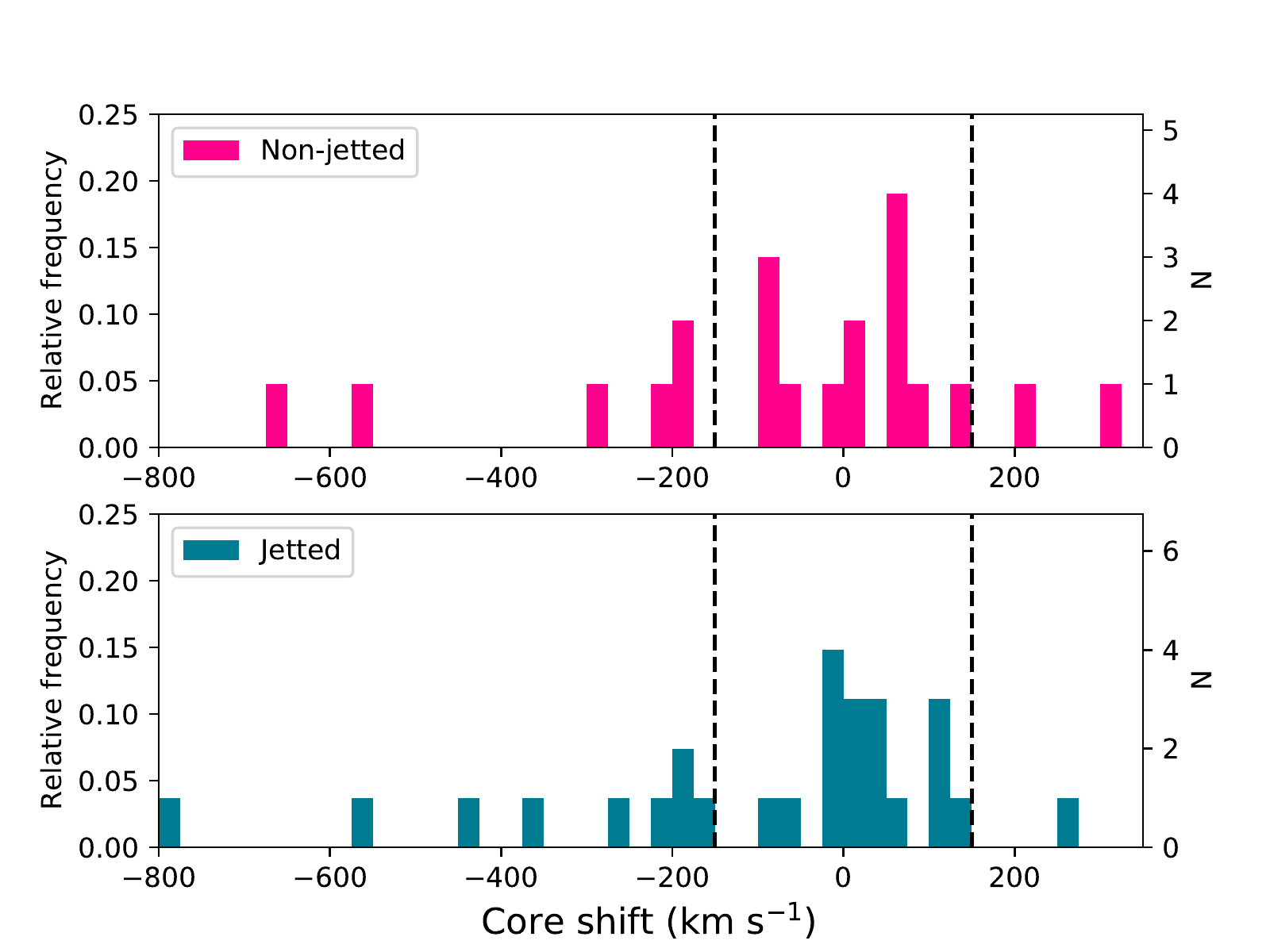}
    \end{minipage}
\hfill
    \begin{minipage}[t]{.49\textwidth}
        \centering
	\includegraphics[trim={0cm 0cm 0cm 0cm}, width=\textwidth]{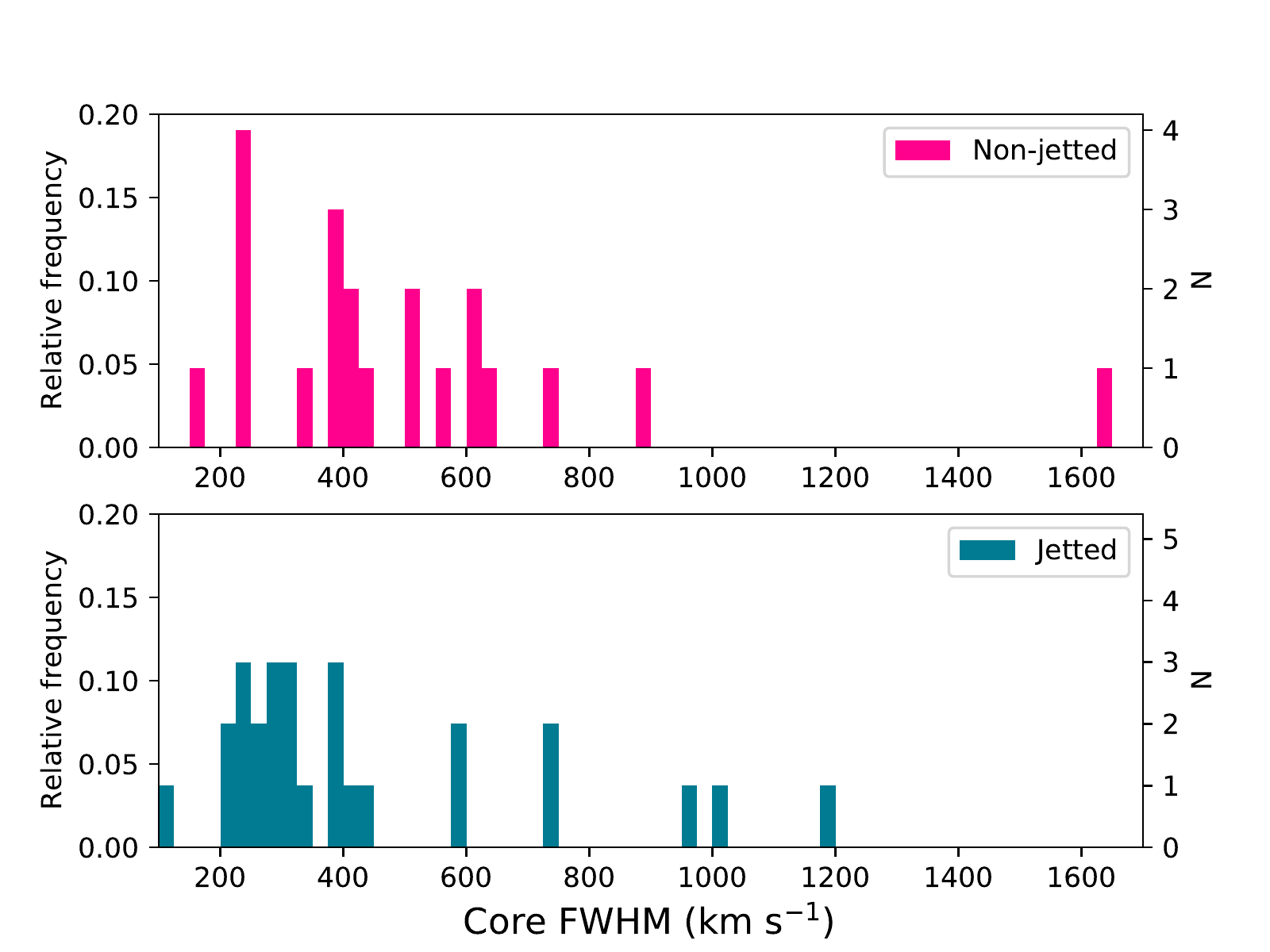}
	\end{minipage}
\caption{\textbf{Left:} distributions of the [O~III] core velocity shift in \kms, in steep-spectrum (top panel) and flat-spectrum jetted sources (bottom panel). The vertical dashed line indicate the limit for red or blue outliers at 150 \kms. Binning of 25 \kms. \textbf{Right:} distributions of the FWHM([O~III]) wing component in \kms\ in the same samples. Binning of 25 \kms.}
\label{fig:spind-core}
\end{figure*}\noindent

The resulting distributions of jetted NLS1s are shown in the left panel of Fig.~\ref{fig:spind-core}. The number of blue and red outliers in the two samples is essentially the same. Out of 21 steep-spectrum sources, we found two red outliers and six blue, while out of 27 flat-spectrum sources we found one red outlier and nine blue. Neither the K-S nor the A-D test allows us to reject the null hypothesis that the two distributions of core shift velocities originate from the same sample of sources. The p-value is always above 0.9, both in our samples and in the simulated ones, indicating that the two distributions are very similar. Also median velocity, standard deviation and IQR are comparable among the two samples. We also analyzed the distributions of FHWM of the [O~III] core component, shown on the right side of Fig.~\ref{fig:spind-core}. As for the core shift distributions, the null hypothesis can never be rejected. In a similar way, the properties of the [O~III] wings among the two samples, both their velocity and FWHM, show that they originate from the same population of sources. All the values we measured are shown in Table~\ref{tab:summary_spind}. \\
It is worth noting that we decided to include the seven sources with absorbed radio jets in the flat-spectrum sample even if, formally, at low radio frequency they either show a steep spectrum, or they are not detected. The reason for this is that their rapid variability observed at high radio frequency indicates that they harbor relativistic jets aligned with the line of sight, as seen in blazars. In fact, their spectral indexes are steep because the radio emission at low frequency is dominated by star formation, but their intrinsic spectrum, if corrected for absorption, has to be flat. However, our conclusions are independent of this assumption. Even when all of them are included in the steep-spectrum sample, the statistical tests still do not provide any evidence against the null hypothesis. This indicates that our conclusion is rather robust, and that the radio spectral index is not directly connected to the physical properties of the [O~III] line. \\

\begin{table}[t]
\caption{Results of the statistical tests for the spectral index samples. Columns: (1) Tested sample; (2) median velocity; (3) standard deviation of the velocity; (4) interquartile range; (5) p-value of the K-S tests against the other sample, observed in top row, simulated in middle row, and standard deviation of the simulated p-value in bottom row; (6) p-value of the A-D tests against the other sample, observed in top row, simulated in middle row, and standard deviation of the simulated p-value in bottom row.}
\centering
\begin{tabular}{l|ccc|ccccccccccccc}
\hline
Sample & \makecell{Median \\ (\kms) } & \makecell{Std \\ (\kms) } & \makecell{IQR \\ (\kms) } & \makecell{K-S$^{obs}$ \\ K-S$^{sim}$ \\ K-S$^{std}$} & \makecell{A-D$^{obs}$ \\ A-D$^{sim}$ \\ A-D$^{std}$} \\
\hline
v$_c$(S) & -4.97 & 226.6 & 252.1 & \makecell{0.71 \\ 0.92 \\ 0.14} & \makecell{0.97 \\ 0.95 \\ 0.11} \\
\hline
v$_c$(F) & -12.8 & 225.8 & 226.2 & \makecell{0.71 \\ 0.92 \\ 0.14} & \makecell{0.97 \\ 0.95 \\ 0.11} \\
\hline
FWHM$_c$(S) & 408.9 & 311.6 & 266.5 & \makecell{0.15 \\ 0.14 \\ 0.11} & \makecell{0.25 \\ 0.28 \\ 0.09} \\
\hline
FWHM$_c$(F) & 321.1 & 268.0 & 253.5 & \makecell{0.15 \\ 0.14 \\ 0.11} & \makecell{0.25 \\ 0.28 \\ 0.09} \\
\hline
v$_w$(S) & -362.1 & 246.5 & 355.6 & \makecell{0.50 \\ 0.50 \\ 0.23} & \makecell{0.26 \\ 0.32 \\ 0.23} \\
\hline
v$_w$(F) & -326.1 & 229.6 & 339.9 & \makecell{0.50 \\ 0.50 \\ 0.23} & \makecell{0.26 \\ 0.32 \\ 0.23} \\
\hline
FWHM$_w$(S) & 1062.6 & 399.4 & 678.3 & \makecell{0.68 \\ 0.68 \\ 0.05} & \makecell{0.67 \\ 0.69 \\ 0.10} \\
\hline
FWHM$_w$(F) & 1180.3 & 445.3 & 467.8 & \makecell{0.68 \\ 0.68 \\ 0.05} & \makecell{0.67 \\ 0.69 \\ 0.10} \\
\hline
\end{tabular}
\label{tab:summary_spind}
\end{table}

\section{Discussion}
\subsection{Origin of the bulk outflows}
The [O~III] lines of our sample could be modeled with two components in 67 out of 77 sources in which the line is detected (87\%), a fraction which is in agreement with previous findings in the literature \citep[e.g.,][]{Mullaney13, Cracco16}. In type 2 and intermediate type AGN, as well as in star forming galaxies, instead, the fraction of two-components [O~III] lines is significantly lower \citep{Vaona12, Concas17}. This indicates that in type 1 AGN two separate kinematic components are directly visible. The first one is the [O~III] core, which is usually not moving with respect to the galaxy restframe, and the second one is the wing, associated to an outflowing and more turbulent component closer to the central engine \citep{Bian05}. \\
In some cases both the [O~III] core and wing components are shifted toward bluer wavelengths, although outflows in the opposite direction can also be observed, likely depending on the geometry of the source. The presence of such wavelength shift is interpreted as a sign of a bulk outflowing motion in their NLR. The existence of these outflows in the NLR of AGN has been known for a long time, and it is particularly common in NLS1s and other population A sources \citep{Boroson92, Zamanov02, Negrete18}. In particular, \citet{Ganci19}, analyzing high-quality optical spectra, found that line displacements are often found in sources with prominent Fe~II multiplets, a feature which is typically observed in NLS1s. This spectral property has often been associated to the high Eddington ratio that is characteristic of population A sources, but past studies showed that these two quantities are not directly correlated \citep[][B16]{Aoki05, Cracco16}. An alternative explanation is that the origin of this phenomenon resides in the relativistic jets crossing the ISM and interacting with it \citep{Komossa08}. Such phenomenon, indeed, has been observed not only in NLS1s, but in several galaxies harboring compact jets \citep{Venturi20}. \\
Our results seem to support the latter scenario. As seen in the left panel of Fig.~\ref{fig:blue_outliers}, the number of blue (and red) outliers observed in sources with relativistic jets is significantly higher than what is found in the non-jetted sample. This finding is also in agreement with the main result of B16. In that work, however, the adopted radio classification was based on the radio-loudness criterion, which has been repeatedly proven to be rather ineffective \citep{Caccianiga15, Padovani17, Jarvela17, Lahteenmaki18, Berton20b, Chen20, Foschini20}. In this work, instead, we make use of a more rigorous classification based directly on the observed radio properties and morphology of the sources. The relativistic jets of NLS1s are usually rather compact \citep{Gu15, Berton18c, Liao20, Chen20}, although some exceptions are known \citep[e.g.,][]{Gliozzi10, Richards15, Congiu17, Rakshit18, Gabanyi18, Congiu20}. This has led some authors to suggest that NLS1s may be connected with classes of young radio galaxies such as compact steep-spectrum (CSS) sources and gigahertz-peaked sources (GPS) \citep{Oshlack01, Gallo06, Komossa08, Caccianiga14, Berton17, Caccianiga17, Liao20, Odea20, Yao20}. Indeed, several radio galaxies show signs of line shifts and broadening induced by the presence of jets \citep{Tadhunter01, Marziani03, Holt06, Venturi20}, and this seems particularly common in young radio galaxies, where the jet is directly interacting with the ISM \citep{Morganti15, Orienti16}. In this framework, the detection of a large fraction of outliers among jetted NLS1s is not surprising. \\
As previously mentioned, \citet{Ganci19} found that line displacements are commonly, although not exclusively, present in sources with prominent Fe~II emission. Our data do not allow to test this conclusion, because accurate Fe~II measurements would require high signal-to-noise (S/N) ratio spectra which are currently unavailable. On the other hand, their sources are not prominent radio emitters according to the FIRST and NVSS radio surveys at 1.4~GHz \citep{Becker95, Condon98}. Potentially, this could indicate that they do not harbor relativistic jets, and that two different physical mechanisms causing outflows can (co-)exist, a first one connected to the relativistic jet, and a second one connected to the presence of iron multiplets. In particular, the iron abundance may be connected to a high Eddington ratio, as postulated by the quasar main sequence \citep{Marziani18b}. This property can produce strong outflows that can ultimately produce the NLR bulk motion we observed. Indeed, \citet{Marziani16b} observed that at intermediate redshift (0.9 $<$ z $<$ 2.2), blue outliers seem to be significantly more common than what is seen at low redshifts, with a whopping detection rate of $\sim$40\%. Those quasars, however, have bolometric luminosity in the range 47 $< \log{L_{\rm Bol}} <$ 48.5 \ergs, that is one order of magnitude more than the brightest source included in our sample. We therefore speculate that, at such extreme luminosity, radiation pressure driven outflows are the most likely mechanism to produce blue outliers. This, of course, preferably happens at relatively high redshift, since in the local Universe such extremely bright objects are usually not observed. When, instead, the bolometric luminosity is smaller, it is possible that the formation of blue outliers is driven mostly by the presence of relativistic jets, although the radiation pressure channel cannot be fully ruled out, since we do observe outliers among non-jetted AGN. \\
It is worth noting that a weak radio emission at low frequencies does not necessarily rule out the presence of relativistic jets. As found by \citet{Berton20b}, the radio emission of some small-scale jets can be entirely absorbed via free-free or synchrotron self-absorption, thus making these features invisible in the low-frequency domain ($<$10~GHz). Their interaction with the ISM is instead well visible, since two out of seven of these peculiar objects have shifted [O~III] lines (J1232+4957, J1509+6137). Therefore, we suggest that blue outliers, especially those showing the most prominent line displacements, could be potential targets in which an absorbed jet could be present. Only high-frequency radio observations will allow us to clarify this aspect. \\
As in \citet{Komossa18a}, we found a rather strong correlation between the FWHM and the shift of the [O~III] core component (Fig.~\ref{fig:core-fwhm}). This seem to indicate that a bulk motion in the NLR (the core shift) is characterized by a larger intrinsic turbulence of the gas (the [O~III] core FHWM). This result was observed not only in [O~III] lines, but also in other NLR lines (such as [Fe~VII], [Ne~III], [Ne~V], [Fe~X]), and in the BLR line C~IV $\lambda$1549 \citep{Komossa08, Spoon09, Gaskell13, Cracco16}. In our analysis, this correlation is dominated by the outliers found in the jetted sources, again suggesting that this behavior can be produced by the relativistic jets crossing the NLR. Another effect of this interaction may be seen in the wing component, since its velocity distribution is different in jetted and non-jetted sources, with the former having a larger scatter than the latter. \\
Another interesting result of our analysis is the non-detection of the [O~III] line in three of our sources, J0744+5149, J0945+1915, and J1421+2824. Except for J0744+5149, where the S/N ratio in the continuum is low ($\sim3$), the other sources have relatively good spectra (S/N$>$10). Therefore, the non-detection in these cases is not due to a poor data quality, but to an intrinsic weakness of the [O~III] lines. All of these sources have relativistic jets with a compact morphology and a flat radio spectrum. This may not be a coincidence. The presence of relativistic jets can decrease the covering factor of the clouds where high ionization lines such as [O~III] originate, and as a consequence reduce the equivalent width and luminosity of the lines \citep{Cecil02, Ludwig12}. Therefore, the non detection of these lines may be another sign of jet/NLR interaction. \\ 
The reason why all jetted sources are not blue outliers is still not fully understood. A possible explanation for this behavior, however, was provided in the simulations by \citet{Wagner12}. The interaction between the relativistic jets and the NLR can indeed occur only if the ratio between the total jet power and the Eddington luminosity of the source exceeds a threshold value (log(P$_j$/L$_{Edd}$) $>$ -4). When this condition is met, the velocity of the outflow produced by the relativistic jet of power P$_j$ exceeds the gas velocity dispersion derived from the relation between the black hole mass and the stellar velocity dispersion \citep{Ferrarese00} of a galaxy for which the Eddington luminosity is L$_{Edd}$. Therefore, the gas can escape the gravitational potential of the black hole. Unfortunately, our data do not allow us to test this hypothesis for all of our sources. However, for a few sources the jet power was calculated by \citet{Foschini15}, and all of them respect this condition, as already shown by B16. 
In conclusion, we speculate that only the high-power jets launched by relatively low-mass - and hence low Eddington luminosity - sources, such as in NLS1s, can produce blue outliers.

\subsection{[O~III] and radio properties}
A direct comparison between the radio morphology and the [O~III] properties in high Eddington ratio AGN such as NLS1s has never been carried out, since the number of radio surveys aimed at studying the kpc-scale properties of these objects is very limited. The role of the radio morphology was studied by dividing the sources in compact, intermediate, and extended, based on the concentration of the flux density in the radio core. As found by B18, jetted sources often have compact or intermediate morphology, while non-jetted ones are mostly extended. However, our results seem to indicate that this is not connected to any of the [O~III] properties we measured. The number of outliers found in the different morphological samples is basically the same, and also the other properties do not show preferences for any sample. This is somewhat surprising taking into account that jetted sources preferably have a compact morphology, and we therefore would have expected to see this trend reflected on the morphological distributions. \\
Interestingly, we also found that the [O~III] properties in jetted sources are the same when the radio spectral index is flat or steep. As previously described, the spectral index can be used as a rough proxy to estimate the source inclination. Flat-spectrum objects are preferentially observed at low inclination with respect to the line of sight, while in steep-spectrum sources the relativistic jets are observed at larger angles. This suggests that the perturbations in the ionized gas kinematics can be seen regardless of the jet inclination. Since the structure of AGN has a cylindrical symmetry, also this result is slightly counterintuitive, as one would expect to observe outflows preferentially at low inclination \citep{Komossa06, Yuan08}, but it is not utterly surprising. If orientation were the sole responsible for blue outliers, we should see them in sources with the highest relativistic beaming and, as a consequence, the highest radio luminosity. However, this has never been the case, as clearly shown in the past \citep{Berton16b, Komossa18a}. Our work confirms this result. The Pearson correlation coefficient between the core shift velocity and the total radio luminosity at 5~GHz is 0.005, with a p-value of 0.97, providing no evidence that these two quantities are connected. Also no correlation was found in the jetted and non-jetted subsamples (see Table~\ref{tab:correlations}). \\
Determining the exact inclination of an AGN, especially non-jetted, is not an easy task. The most reliable technique relies on the geometry of the ionization cones, but it can be applied only for a limited number of sources \citep{Marin16}. Blue outliers, until now, were commonly considered low-inclination sources. Indeed, \citet{Boroson11} suggested that they could be used, along with the red-shift of Fe~II multiplets (\citep{Hu08a}, but see \citep{Bon18}), as inclination indicators. However, our findings seriously challenge this conclusion, since blue outliers seem to be uniformly distributed among sources with different inclinations. \\
The cloud acceleration process, therefore, seems to occur in every direction, and not only along the jet propagation axis. This behavior is similar to what was predicted by hydrodynamical simulations by \citet{Wagner11} and \citet{Wagner12}\footnote{A very explanatory video can be found at https://www.youtube.com/watch?v=6-CO4hp6PYg}. In their simulations, the jet can accelerate the clouds to velocities up to $10^3$ \kms, which are comparable to the highest values we observed in our sample. Similarly, \citet{Cielo18} found that young jets can accelerate the gas clouds up to these velocities much more efficiently than what nuclear outflows induced by disk winds can do. Therefore, we can conclude that faster blue outliers are the most likely to originate from jet-ISM interaction. \\

\section{Summary}
In this paper, we analyzed the properties of the [O~III] lines in a sample of low-mass high Eddington ratio AGN belonging to the NLS1 class, and compared them with their radio properties reported by a recent JVLA survey at 5~GHz. We found strong evidence that the presence of relativistic jets in these AGN can produce a bulk outflowing motion in the NLR, causing the [O~III] lines to appear blue-shifted with respect to their restframe wavelength. Among the non-jetted AGN we found one outlier out of 28 sources, while amidst the jetted AGN we identified 18 outliers out of 48 sources. No strong correlation between the bolometric luminosity of the sources and the [O~III] line properties was found, suggesting only a minor contribution of radiation pressure driven outflows in producing [O~III] shifts. We therefore suggest that the most common origin for these so-called blue outliers is the interaction between the NLR gas clouds producing [O~III] lines and the relativistic jets. The presence of blue outliers seems not to be connected either with the radio morphology of the source, or with the inclination of radio jets. This possibly indicates that the acceleration of the clouds produced by the propagation of the jets in the NLR medium does not have a preferred direction. The association between the blue outliers and the relativistic jets might indicate that the properties of the [O~III] lines can be used as a powerful tool to identify the presence of jets whose low-frequency radio emission is absorbed, and that have gone undetected in the past radio surveys. Further studies, especially by means of spatially resolved integral-field spectroscopy, are needed to fully understand the interplay between the relativistic jets and the NLR clouds. 

\bibliographystyle{aa}
\bibliography{./biblio}
\clearpage

\begin{table}[t]
\caption{Results of the correlations tested throughout this work. Columns: (1) First tested sample; (2) second tested sample; (3) Pearson correlation coefficient; (4) p-value.}
\label{tab:correlations}
\centering
\begin{tabular}{lccc}
\hline
Q1 & Q2 & r & p-value \\
\hline
v$_c^{tot}$ & FWHM$_c^{tot}$ & -0.70 & 8$\times10^{-13}$ \\
v$_c^{NJ}$ & FWHM$_c^{NJ}$ & 0.02 & 0.93 \\
v$_c^{J}$ & FWHM$_c^{J}$ & -0.73 & 4$\times10^{-9}$ \\
v$_w^{tot}$ & v$_c^{tot}$ & 0.01 & 0.92 \\
v$_w^{NJ}$ & v$_c^{NJ}$ & 0.20 & 0.35 \\
v$_w^{J}$ & v$_c^{J}$ & 0.24 & 0.13 \\
v$_w^{tot}$ & FWHM$_c^{tot}$ & -0.09 & 0.43 \\
v$_w^{NJ}$ & FWHM$_c^{NJ}$ & -0.06 & 0.79 \\
v$_w^{J}$ & FWHM$_c^{J}$ & -0.50 & 8$\times10^{-4}$ \\
FWHM$_w^{tot}$ & v$_c^{tot}$ & 0.02 & 0.85 \\
FWHM$_w^{NJ}$ & v$_c^{NJ}$ & -0.21 & 0.32 \\
FWHM$_w^{J}$ & v$_c^{J}$ & -0.36 & 0.02 \\
FWHM$_w^{tot}$ & FWHM$_c^{tot}$ & -0.01 & 0.95 \\
FWHM$_w^{NJ}$ & FWHM$_c^{NJ}$ & 0.34 & 0.09 \\
FWHM$_w^{J}$ & FWHM$_c^{J}$ & 0.55 & 2$\times10^{-4}$ \\
v$_w^{tot}$ & FWHM$_w^{tot}$ & -0.47 & 1$\times10^{-5}$ \\
v$_w^{NJ}$ & FWHM$_w^{NJ}$ & -0.51 & 0.01 \\
v$_w^{J}$ & FWHM$_w^{J}$ & -0.08 & 0.61 \\
v$_c^{NJ}$ & L$^{5~GHz}_{int}$ & -0.30 & 0.12 \\
v$_c^{J}$ & L$^{5~GHz}_{int}$ & 0.03 & 0.84 \\
v$_c^{NJ}$ & L$_{[O~III]}$ & 0.22 & 0.26 \\
v$_c^{J}$ & L$_{[O~III]}$ & -0.01 & 0.94 \\

\hline
\end{tabular}
\end{table}

\clearpage 
\footnotesize

\begin{longtable}{l c c c c c c c}
\caption{Sample studied in this work. Columns: (1) short name; (2) NED alias; (3) right ascension; (4) declination; (5) redshift; (6) jet (Y stands for yes, N for no); (7) radio morphology (C stands for compact, I for intermediate, E for extended); (8) spectral index for jetted sources (F stands for flat spectrum, S stands for steep spectrum).}
\label{tab:sample} \\
\hline 
\multicolumn{1}{c}{Short Name} & \multicolumn{1}{c}{NED Alias} & \multicolumn{1}{c}{R.A.} & \multicolumn{1}{c}{Dec.} & \multicolumn{1}{c}{z} & \multicolumn{1}{c}{Jet} & \multicolumn{1}{c}{Morph} & \multicolumn{1}{c}{Sp.Ind.} \\ \hline 
\endfirsthead
\multicolumn{8}{c}%
{{\it \tablename\ \thetable{} -- continued from previous page}} \\
\hline \multicolumn{1}{c}{Short Name} & \multicolumn{1}{c}{NED Alias} & \multicolumn{1}{c}{R.A.} & \multicolumn{1}{c}{Dec.} & \multicolumn{1}{c}{z} & \multicolumn{1}{c}{Jet} & \multicolumn{1}{c}{Morph} & \multicolumn{1}{c}{Sp.Ind.} \\
\hline 
\endhead
\hline \multicolumn{8}{r}{\textit{Continued on next page}} \\ \hline
\endfoot
\hline \hline
\endlastfoot
J0006$+$2012 & Mrk 335 & 00 06 19.50 & +20 12 10.0 & 0.026 & N & C & $-$ \\
J0146$-$0040 & 2MASX J01464481$-$0040426 & 01 46 44.80 & $-$00 40 43.0 & 0.083 & N & C & $-$ \\
J0347$+$0105 & IRAS 03450+0055 & 03 47 40.20 & +01 05 14.0 & 0.031 & N & E & $-$ \\ 
J0629$-$0545 & IRAS 06269$-$0543 & 06 29 24.70 & $-$05 45 30.0 & 0.117 & Y & I & S \\
J0632$+$6340 & UGC 3478 & 06 32 47.20 & +63 40 25.0 & 0.013 & N & E & $-$ \\
J0706$+$3901 & FBQS J0706+3901 & 07 06 25.15 & +39 01 51.6 & 0.086 & Y & I & S \\
J0713$+$3820 & FBQS J0713+3820 & 07 13 40.29 & +38 20 40.1 & 0.123 & Y & E & S \\
J0744$+$5149 & NVSS J074402+514917 & 07 44 02.24 & +51 49 17.5 & 0.460 & Y & C & F \\
J0752$+$2617 & FBQS J0752+2617 & 07 52 45.60 & +26 17 36.0 & 0.082 & N & C & $-$ \\
J0758$+$3920 & FBQS J075800.0+392029 & 07 58 00.05 & +39 20 29.1 & 0.096 & Y & E & S \\
J0804$+$3853 & FBQS J0804+3853 & 08 04 09.24 & +38 53 48.7 & 0.212 & Y & E & S \\ 
J0806$+$7248 & RGB J0806+728 & 08 06 38.96 & +72 48 20.4 & 0.098 & Y & I & S \\ 
J0814$+$5609 & SDSS J081432.11+560956.6 & 08 14 32.13 & +56 09 56.6 & 0.510 & Y & I & F \\
J0849$+$5108 & SBS 0846+513 & 08 49 57.99 & +51 08 28.8 & 0.585 & Y & C & F \\
J0850$+$4626 & SDSS J085001.17+462600.5 & 08 50 01.17 & +46 26 00.5 & 0.524 & Y & C & F \\
J0902$+$0443 & SDSS J090227.16+044309.5 & 09 02 27.15 & +04 43 09.4 & 0.533 & Y & C & F \\
J0913$+$3658 & RX J0913.2+3658 & 09 13 13.70 & +36 58 17.0 & 0.107 & N & I & $-$ \\
J0925$+$5217 & Mrk 110 & 09 25 12.90 & +52 17 11.0 & 0.035 & N & E & $-$ \\
J0926$+$1244 & Mrk 705 & 09 26 03.30 & +12 44 04.0 & 0.029 & N & I & $-$ \\
J0937$+$3615 & SDSS J093703.03+361537.2 & 09 37 03.01 & +36 15 37.3 & 0.180 & N & E & $-$ \\
J0945$+$1915 & SDSS J094529.23+191548.8 & 09 45 29.21 & +19 15 48.9 & 0.284 & Y & C & F \\
J0948$+$5029 & Mrk 124 & 09 48 42.60 & +50 29 31.0 & 0.056 & N & C & $-$ \\
J0952$-$0136 & Mrk 1239 & 09 52 19.10 & $-$01 36 43.0 & 0.020 & Y & I & S \\
J0957$+$2433 & RX J0957.1+2433 & 09 57 07.20 & +24 33 16.0 & 0.082 & N & C & $-$ \\
J1029$+$5556 & SDSS J102906.69+555625.2 & 10 29 06.69 & +55 56 25.2 & 0.451 & Y & C & F \\
J1031$+$4234 & SDSS J103123.73+423439.3 & 10 31 23.73 & +42 34 39.4 & 0.377 & Y & C & S \\
J1034$+$3938 & KUG 1031+398 & 10 34 38.60 & +39 38 28.0 & 0.042 & Y & I & S \\
J1037$+$0036 & SDSS J103727.45+003635.6 & 10 37 27.45 & +00 36 35.8 & 0.595 & Y & C & F \\
J1038$+$4227 & SDSS J103859.58+422742.3 & 10 38 59.59 & +42 27 42.0 & 0.220 & N & E & $-$ \\
J1047$+$4725 & SDSS J104732.68+472532.0 & 10 47 32.65 & +47 25 32.2 & 0.799 & Y & E & F \\
J1048$+$2222 & SDSS J104816.57+222238.9 & 10 48 16.56 & +22 22 40.1 & 0.330 & Y & I & S \\
J1102$+$2239 & SDSS J110223.38+223920.7 & 11 02 23.36 & +22 39 20.7 & 0.453 & Y & I & S \\
J1110$+$3653 & SDSS J111005.03+365336.3 & 11 10 05.03 & +36 53 36.1 & 0.629 & Y & I & F \\
J1114$+$3241 & B2 1111+32 & 11 14 38.89 & +32 41 33.4 & 0.189 & Y & C & F \\
J1121$+$5351 & SBS 1118+541 & 11 21 08.60 & +53 51 21.0 & 0.103 & N & I & $-$ \\
J1138$+$3653 & SDSS J113824.54+365327.1 & 11 38 24.54 & +36 53 27.1 & 0.356 & Y & I & S \\
J1146$+$3236 & SDSS J114654.28+323652.3 & 11 46 54.30 & +32 36 52.2 & 0.465 & Y & C & F \\
J1159$+$2838 & SDSS J115917.32+283814.5 & 11 59 17.31 & +28 38 14.8 & 0.210 & Y & I & S \\
J1203$+$4431 & NGC 4051 & 12 03 09.60 & +44 31 53.0 & 0.002 & N & E & $-$ \\
J1209$+$3217 & RX J1209.7+3217 & 12 09 45.20 & +32 17 01.0 & 0.144 & N & I & $-$ \\
J1215$+$5442 & SBS 1213+549A & 12 15 49.40 & +54 42 24.0 & 0.150 & N & I & $-$ \\
J1218$+$2948 & Mrk 766 & 12 18 26.50 & +29 48 45.8 & 0.013 & N & E & $-$ \\
J1227$+$3214 & SDSS J122749.14+321458.9 & 12 27 49.15 & +32 14 59.0 & 0.136 & Y & I & S \\
J1228$+$5017 & SDSS J122844.81+501751.2 & 12 28 44.81 & +50 17 51.2 & 0.262 & Y & I & F \\
J1232$+$4957 & SDSS J123220.11+495721.8 & 12 32 20.11 & +49 57 21.8 & 0.262 & Y & C & F \\
J1238$+$3942 & SDSS J123852.12+394227.8 & 12 27 49.15 & +39 42 27.6 & 0.622 & Y & C & F \\
J1242$+$3317 & WAS 61 & 12 42 10.60 & +33 17 03.0 & 0.044 & N & E & $-$ \\
J1246$+$0222 & PG 1244+026 & 12 46 35.20 & +02 22 09.0 & 0.048 & N & E & $-$ \\
J1246$+$0238 & SDSS J124634.65+023809.0 & 12 46 34.68 & +02 38 09.0 & 0.367 & Y & C & F \\
J1302$+$1624 & Mrk 783 & 13 02 58.80 & +16 24 27.0 & 0.067 & Y & E & S \\
J1305$+$5116 & SDSS J130522.74+511640.2 & 13 05 22.75 & +51 16 40.3 & 0.788 & Y & E & F \\
J1317$+$6010 & SBS 1315+604 & 13 17 50.30 & +60 10 41.0 & 0.137 & N & E & $-$ \\
J1333$+$4141 & SDSS J133345.47+414127.7 & 13 33 45.47 & +41 41 28.2 & 0.225 & Y & C & S \\
J1337$+$2423 & IRAS 13349+2438 & 13 37 18.70 & +24 23 03.0 & 0.108 & N & I & $-$ \\
J1346$+$3121 & SDSS J134634.97+312133.7 & 13 46 35.07 & +31 21 33.9 & 0.246 & Y & C & F \\
J1355$+$5612 & SBS 1353+564 & 13 55 16.50 & +56 12 45.0 & 0.122 & N & C & $-$ \\
J1358$+$2658 & SDSS J135845.38+265808.4 & 13 58 45.40 & +26 58 08.3 & 0.331 & Y & I & S \\
J1402$+$2159 & RX J1402.5+2159 & 14 02 34.40 & +21 59 52.0 & 0.066 & N & I & $-$ \\
J1421$+$2824 & SDSS J142114.05+282452.8 & 14 21 14.07 & +28 24 52.2 & 0.539 & Y & C & F \\
J1443$+$4725 & SDSS J144318.56+472556.7 & 14 43 18.56 & +47 25 56.7 & 0.705 & Y & C & F \\
J1505$+$0326 & SDSS J150506.47+032630.8 & 15 05 06.47 & +03 26 30.8 & 0.408 & Y & C & F \\
J1509$+$6137 & SDSS J150916.18+613716.7 & 15 09 16.18 & +61 37 16.7 & 0.201 & Y & C & F \\
J1510$+$5547 & SDSS J151020.06+554722.0 & 15 10 20.06 & +55 47 22.0 & 0.150 & Y & E & F \\
J1522$+$3934 & SDSS J152205.41+393441.3 & 15 22 05.41 & +39 34 41.3 & 0.077 & Y & I & F \\
J1536$+$5433 & Mrk 486 & 15 36 38.30 & +54 33 33.0 & 0.039 & N & I & $-$ \\
J1537$+$4942 & SDSS J153732.61+494247.5 & 15 37 32.60 & +49 42 48.0 & 0.280 & N & C & $-$ \\
J1548$+$3511 & SDSS J154817.92+351128.0 & 15 48 17.92 & +35 11 28.4 & 0.479 & Y & C & F \\
J1555$+$1911 & Mrk 291 & 15 55 07.90 & +19 11 33.0 & 0.035 & N & E & $-$ \\
J1559$+$3501 & Mrk 493 & 15 59 09.60 & +35 01 47.0 & 0.031 & N & E & $-$ \\
J1612$+$4219 & SDSS J161259.83+421940.3 & 16 12 59.83 & +42 19 40.0 & 0.233 & Y & E & F \\
J1629$+$4007 & SDSS J162901.30+400759.9 & 16 29 01.31 & +40 07 59.6 & 0.272 & Y & C & F \\
J1633$+$4718 & SDSS J163323.58+471858.9 & 16 33 23.58 & +47 18 59.0 & 0.116 & Y & I & F \\
J1634$+$4809 & SDSS J163401.94+480940.2 & 16 34 01.94 & +48 09 40.1 & 0.495 & Y & C & F \\
J1641$+$3454 & SDSS J164100.10+345452.7 & 16 41 00.10 & +34 54 52.7 & 0.164 & Y & E & F \\
J1703$+$4540 & SDSS J170330.38+454047.1 & 17 03 30.38 & +45 40 47.2 & 0.060 & Y & I & S \\
J1709$+$2348 & SDSS J170907.80+234837.7 & 17 09 07.82 & +23 48 38.2 & 0.254 & Y & C & S \\
J1713$+$3523 & FBQS J1713+3523 & 17 13 04.48 & +35 23 33.4 & 0.083 & Y & I & S \\
J2242$+$2943 & Ark 564 & 22 42 39.30 & +29 43 31.0 & 0.025 & N & E & $-$ \\
J2314$+$2243 & RX J2314.9+2243 & 23 14 55.70 & +22 43 25.0 & 0.169 & Y & I & S \\
\end{longtable} 

\clearpage
\begin{landscape}
\begin{longtable}{l c c c c c c c c c c}
\caption{[O~III] and radio properties of the sample. Columns: (1) short name; (2) logarithm of the [O~III] luminosity (\ergs); (3) wavelength of the [O~III] core component (\AA); (4) velocity of the [O~III] core component (\kms); (5) FWHM of the [O~III] core component (\kms); (6) wavelength of the [O~III] wing component (\AA); (7) velocity of the [O~III] wing component (\kms); (8) FWHM of the [O~III] wing component (\kms); (9) logarithm of the integrated radio luminosity at 5~GHz (\ergs); (10) logarithm of the peak luminosity at 5~GHz (\ergs); (11) logarithm of the diffuse luminosity at 5~GHz (\ergs).}
\label{tab:oiii} \\
\hline 
\multicolumn{1}{c}{Short Name} & \multicolumn{1}{c}{$\log L_{[O~III]}$} & \multicolumn{1}{c}{$\lambda_c$} & \multicolumn{1}{c}{v$_c$} & \multicolumn{1}{c}{FWHM$_c$} & \multicolumn{1}{c}{$\lambda_w$} & \multicolumn{1}{c}{v$_w$} & \multicolumn{1}{c}{FWHM$_w$} & \multicolumn{1}{c}{$\log L_{int}$} & \multicolumn{1}{c}{$\log L_{peak}$} & \multicolumn{1}{c}{$\log L_{diff}$} \\ \hline 
\endfirsthead
\multicolumn{11}{c}
{{\it \tablename\ \thetable{} -- continued from previous page}} \\
\hline \multicolumn{1}{c}{Short Name} & \multicolumn{1}{c}{$\log L_{[O~III]}$} & \multicolumn{1}{c}{$\lambda_c$} & \multicolumn{1}{c}{v$_c$} & \multicolumn{1}{c}{FWHM$_c$} & \multicolumn{1}{c}{$\lambda_w$} & \multicolumn{1}{c}{v$_w$} & \multicolumn{1}{c}{FWHM$_w$} & \multicolumn{1}{c}{$\log L_{int}$} & \multicolumn{1}{c}{$\log L_{peak}$} & \multicolumn{1}{c}{$\log L_{diff}$} \\ 
\hline 
\endhead
\hline \multicolumn{11}{r}{\textit{Continued on next page}} \\ \hline
\endfoot
\hline \hline
\endlastfoot
J0006+2012  & 41.47 & 5005.65$\pm$0.02 &  -71.63$\pm$1.03 &  458.28$\pm$4.77 & 5001.29$\pm$0.11 &  -332.20$\pm$6.84 & 1387.51$\pm$17.02 & 38.42 & 38.41 & 36.88 \\ 
J0146-0040  & 41.16 & 5006.67$\pm$0.05 &  -10.36$\pm$2.99 &  143.83$\pm$4.19 & 5005.78$\pm$0.22 &   -53.34$\pm$14.37 &  575.32$\pm$19.16 & 37.95 & 37.94 & 36.40 \\ 
J0347+0105  & 41.32 & 5004.23$\pm$0.12 & -156.55$\pm$7.33 &  328.92$\pm$26.55 & 5000.64$\pm$0.15 &  -371.31$\pm$9.11 & 1074.27$\pm$27.42 & 39.17 & 39.01 & 38.66 \\ 
J0629-0543  & 42.63 & 5007.91$\pm$0.25 &   63.91$\pm$15.15 &  621.47$\pm$60.71 & 4999.09$\pm$0.92 &  -463.97$\pm$55.21 & 1627.96$\pm$52.20 & 40.43 & 40.36 & 39.59 \\ 
J0632+6340  & 40.28 & 5007.03$\pm$0.22 &   11.20$\pm$13.17 &  547.12$\pm$4.19 & 5002.18$\pm$3.92 &  -290.47$\pm$242.50 &  999.76$\pm$115.56 & 37.76 & 37.54 & 37.39 \\ 
J0706+3901  & 41.33 & 5003.61$\pm$0.07 & -193.58$\pm$4.19 &  736.07$\pm$2.99 &    0.00 &       0.00 &     0.00 & 39.33 & 39.23 & 38.63 \\ 
J0713+3820  & 42.39 & 5011.86$\pm$0.14 &   300.4$\pm$8.38 &  504.81$\pm$12.57 & 4999.46$\pm$0.58 &  -741.81$\pm$41.31 & 1494.70$\pm$17.96 & 39.82 & 39.68 & 39.27 \\ 
J0744+5149 & N.D. &    &            &         &             &            &          & 40.98 & 40.97 & 39.53 \\ 
J0752+2617  & 41.00 & 5006.13$\pm$0.06 &  -42.69$\pm$3.59 &  227.02$\pm$2.40 & 5001.64$\pm$0.31 &  -268.66$\pm$20.36 &  720.56$\pm$8.98 & 39.19 & 38.17 & 36.64 \\ 
J0758+3920  & 42.67 & 5005.40$\pm$0.07 &  -86.40$\pm$4.19 &  644.41$\pm$2.99 & 4993.81$\pm$0.40 &  -694.39$\pm$26.35 &  468.15$\pm$17.96 & 39.68 & 39.39 & 39.36\\ 
J0804+3853  & 42.19 & 5007.01$\pm$0.04 &   10.00$\pm$2.40 &  157.93$\pm$1.20 & 5000.52$\pm$0.15 &  -388.22$\pm$9.58 &  760.04$\pm$3.59 & 39.69 & 39.33 & 39.43 \\ 
J0806+7248  & 41.81 & 5001.90$\pm$0.20 & -295.97$\pm$11.98 &  375.08$\pm$8.98 & 4994.79$\pm$1.49 &  -426.03$\pm$99.40 &  733.25$\pm$61.67 & 40.17 & 40.15 & 38.93 \\ 
J0814+5609  & 41.73 & 4997.51$\pm$0.51 & -558.83$\pm$30.54 &  740.30$\pm$14.97 &         0.00 &    0.00 &         0.00 & 42.19 & 42.13 & 41.28 \\ 
J0849+5108  & 41.74 & 5011.26$\pm$0.18 &  264.47$\pm$10.78 &  231.26$\pm$8.98 & 5002.93$\pm$1.09 &  -498.77$\pm$74.25 &  703.64$\pm$50.30 & 43.21 & 43.20 & 41.22 \\ 
J0850+4626  & 42.32 & 5008.52$\pm$0.16 &  100.41$\pm$9.58 &  248.18$\pm$14.97 & 5005.76$\pm$0.52 &  -164.70$\pm$38.92 & 1558.15$\pm$28.74 & 41.77 & 41.76 & 39.98 \\ 
J0902+0443  & 42.20 & 5007.75$\pm$0.27 &   54.31$\pm$16.17 &  286.25$\pm$26.35 & 4999.03$\pm$0.66 &  -521.95$\pm$54.49 & 1231.01$\pm$20.36 & 42.77 & 42.77 &  0.00 \\ 
J0913+3658  & 41.43 & 5005.90$\pm$0.08 &  -56.46$\pm$4.79 &  329.96$\pm$5.39 & 5000.78$\pm$0.32 &  -306.62$\pm$22.15 &  799.52$\pm$7.78 & 38.66 & 38.61 & 37.74 \\ 
J0925+5217  & 41.99 & 5007.78$\pm$0.53 &    56.1$\pm$31.73 &  215.74$\pm$1.80 &         0.00 &    0.00 &         0.00 & 38.72 & 38.42 & 38.42 \\ 
J0926+1244  & 41.14 & 5006.32$\pm$0.61 &  -31.32$\pm$36.52 &  337.01$\pm$31.73 & 5005.00$\pm$0.96 &   -79.02$\pm$92.21 &  846.06$\pm$32.93 & 38.52 & 38.45 & 37.70 \\ 
J0937+3615  & 41.76 & 5006.98$\pm$0.14 &    8.20$\pm$8.38 &  404.70$\pm$11.98 & 5002.80$\pm$0.69 &  -249.94$\pm$48.50 &  999.76$\pm$25.15 & 39.71 & 39.58 & 39.11 \\ 
J0945+1915 & N.D. &    &           &          &              &         &              & 40.90 & 40.88 & 39.52 \\ 
J0948+5029  & 41.48 & 5006.74$\pm$0.05 &   -6.17$\pm$2.99 &  337.01$\pm$1.80 & 5001.76$\pm$0.17 &  -298.60$\pm$11.38 &  868.62$\pm$5.39 & 38.91 & 38.89 & 37.52 \\ 
J0952-0136  & 41.67 & 5005.96$\pm$0.08 &  -52.87$\pm$4.79 &  558.40$\pm$4.19 & 4994.51$\pm$0.52 &  -685.82$\pm$33.53 & 1583.53$\pm$13.77 & 39.04 & 38.99 & 38.02 \\ 
J0957+2433  & 41.36 & 5005.64$\pm$0.06 &  -72.03$\pm$3.59 &  327.14$\pm$2.40 & 5001.11$\pm$0.27 &  -271.66$\pm$17.96 &  958.86$\pm$9.58 & 38.32 & 38.31 & 36.73 \\ 
J1031+4234  & 41.83 & 5009.07$\pm$0.30 &  133.35$\pm$17.96 &  231.26$\pm$10.78 & 5005.45$\pm$0.92 &  -216.94$\pm$65.86 &  744.53$\pm$10.18 & 41.14 & 41.13 & 39.55 \\ 
J1034+3938  & 41.03 & 5006.76$\pm$0.04 &   -4.97$\pm$2.40 &  228.44$\pm$1.80 & 5002.54$\pm$0.17 &  -253.01$\pm$10.78 &  901.05$\pm$4.79 & 39.24 & 39.18 & 38.34 \\ 
J1037+0036  & 41.60 & 5008.56$\pm$0.42 &  102.81$\pm$25.15 &  280.61$\pm$7.78 & 4999.72$\pm$0.58 &  -529.11$\pm$58.08 &  356.75$\pm$22.15 & 42.19 & 42.19 &  0.00 \\ 
J1038+4227  & 41.62 & 5005.11$\pm$0.12 & -103.77$\pm$7.19 &  314.45$\pm$7.78 & 4999.87$\pm$0.52 &  -313.96$\pm$36.52 & 1146.41$\pm$28.14 & 40.67 & 40.16 & 40.51 \\ 
J1047+4725  & 42.89 & 5006.66$\pm$0.07 &  -10.96$\pm$4.19 &  280.61$\pm$4.79 & 5004.82$\pm$0.39 &  -110.43$\pm$25.75 &  844.65$\pm$24.55 & 43.77 & 43.51 & 43.42 \\ 
J1048+2222  & 41.79 & 5003.35$\pm$0.31 & -209.15$\pm$18.56 &  389.19$\pm$19.76 & 4994.49$\pm$0.75 &  -530.88$\pm$61.07 & 1238.06$\pm$16.77 & 39.75 & 39.72 & 38.59\\ 
J1102+2239  & 42.38 & 4997.40$\pm$0.36 & -565.41$\pm$21.56 &  879.90$\pm$11.38 & 4984.90$\pm$2.41 &  -749.79$\pm$164.06 & 1319.85$\pm$83.23 & 40.47 & 40.44 & 39.35 \\ 
J1110+3653  & 41.73 & 5007.53$\pm$0.22 &   41.14$\pm$13.17 &  400.47$\pm$16.77 &    0.00 &    0.00 &         0.00 & 41.89 & 41.83 & 40.98 \\ 
J1114+3241  & 42.13 & 4993.71$\pm$0.08 & -786.31$\pm$5.06 &  971.78$\pm$22.39 & 4979.31$\pm$0.67 & -1648.58$\pm$40.06 & 2101.97$\pm$36.32 & 41.04 & 41.03 & 39.35\\ 
J1121+5351  & 42.01 & 5007.09$\pm$0.06 &   14.79$\pm$3.59 &  305.99$\pm$2.99 & 5003.61$\pm$0.15 &  -208.07$\pm$10.78 &  734.66$\pm$4.79 & 39.25 & 39.17 & 38.45\\ 
J1138+3653  & 41.68 & 5005.39$\pm$0.17 &  -87.00$\pm$10.18 &  234.08$\pm$16.17 & 5004.59$\pm$0.43 &   -48.40$\pm$34.73 &  923.61$\pm$28.74 & 41.02 & 41.00 & 39.77 \\ 
J1146+3236  & 42.00 & 5007.10$\pm$0.10 &   15.39$\pm$5.99 &  307.40$\pm$4.19 & 5001.47$\pm$0.40 &  -336.86$\pm$28.74 &  620.44$\pm$7.78 & 41.90 & 41.89 & 39.94\\ 
J1159+2838  & 42.36 & 5007.75$\pm$0.04 &   54.32$\pm$2.17 &  397.77$\pm$7.33 & 5002.81$\pm$0.05 &  -241.67$\pm$2.80 & 1201.68$\pm$5.55 & 39.76 & 39.74 & 38.51\\ 
J1203+4431  & 39.94 & 5006.94$\pm$0.45 &    5.81$\pm$26.94 &  318.68$\pm$5.99 & 5002.02$\pm$0.91 &  -294.61$\pm$59.88 &  606.34$\pm$81.43 & 36.46 & 35.93 & 36.31 \\ 
J1209+3217  & 41.74 & 5006.17$\pm$0.19 &  -40.30$\pm$11.38 &  410.34$\pm$6.59 & 5002.46$\pm$0.24 &  -221.92$\pm$23.95 & 1405.86$\pm$12.57 & 39.31 & 39.26 & 38.35 \\ 
J1215+5442  & 41.78 & 5005.69$\pm$0.16 &  -69.04$\pm$9.58 &  384.96$\pm$7.19 & 4997.34$\pm$0.58 &  -500.47$\pm$42.51 &  661.33$\pm$15.57 & 39.30 & 39.25 & 38.34 \\ 
J1218+2948  & 41.26 & 5007.70$\pm$0.38 &   51.31$\pm$22.75 &  341.24$\pm$2.40 & 5005.81$\pm$0.47 &  -113.22$\pm$50.90 &  774.14$\pm$6.59 & 38.47 & 38.34 & 37.87 \\ 
J1227+3214  & 42.11 & 5007.84$\pm$0.05 &   59.70$\pm$2.99 &  242.54$\pm$1.80 & 5004.35$\pm$0.12 &  -209.05$\pm$8.38 &  592.24$\pm$3.59 & 39.94 & 39.90 & 38.92 \\ 
J1238+3942  & 41.98 & 5005.75$\pm$0.15 &  -65.45$\pm$8.98 &  380.73$\pm$15.57 &    0.00 &       0.00 &          0.00 & 42.16 & 42.16 &  0.00 \\ 
J1242+3317  & 41.65 & 5006.86$\pm$0.04 &    1.02$\pm$2.40 &  307.40$\pm$1.20 & 5003.30$\pm$0.10 &  -213.29$\pm$6.59 &  888.36$\pm$2.99 & 38.74 & 38.51 & 38.35 \\ 
J1246+0222  & 41.12 & 5006.03$\pm$0.07 &  -48.68$\pm$4.19 &  310.22$\pm$3.59 & 5002.24$\pm$0.49 &  -227.11$\pm$31.73 &  738.89$\pm$13.77 & 38.31 & 38.12 & 37.87 \\ 
J1246+0238  & 41.64 & 5007.51$\pm$0.35 &   39.94$\pm$20.96 &  382.14$\pm$20.36 & 4999.10$\pm$1.96 &  -503.49$\pm$136.52 & 1245.11$\pm$70.06 & 41.69 & 41.67 & 40.20 \\ 
J1302+1624  & 42.31 & 5003.65$\pm$0.05 & -191.19$\pm$2.99 &  408.93$\pm$1.20 &    0.00 &         0.00 &         0.00 & 39.91 & 39.28 & 39.80 \\ 
J1305+5116  & 43.14 & 5003.62$\pm$1.36 & -192.98$\pm$81.43 & 1012.45$\pm$77.84 & 4993.35$\pm$2.45 &  -615.00$\pm$226.33 & 1394.58$\pm$43.11 & 42.91 & 42.71 & 42.47 \\ 
J1317+6010  & 41.74 & 5007.03$\pm$0.03 &   11.31$\pm$1.55 &  341.35$\pm$6.69 & 5000.55$\pm$0.34 &  -376.82$\pm$20.53 & 1435.93$\pm$33.88 & 39.42 & 39.28 & 38.86 \\ 
J1333+4141  & 42.34 & 5005.35$\pm$0.08 &   -89.4$\pm$4.79 &  521.73$\pm$3.59 & 5001.35$\pm$0.11 &  -239.37$\pm$10.18 & 1745.69$\pm$5.39 & 39.85 & 39.84 & 38.21 \\ 
J1337+2423  & 42.24 & 5005.74$\pm$0.25 &  -66.04$\pm$14.97 &  352.52$\pm$16.17 & 4995.76$\pm$1.29 &  -597.81$\pm$90.41 & 2240.64$\pm$25.75 & 40.26 & 40.23 & 39.18 \\ 
J1346+3121  & 41.39 & 5005.55$\pm$0.51 &  -77.42$\pm$30.54 &  248.18$\pm$55.09 & 5001.72$\pm$2.60 &  -229.05$\pm$184.42 &  920.79$\pm$102.39 & 40.63 & 40.63 &  0.00 \\ 
J1355+5612  & 42.22 & 5007.64$\pm$0.04 &   47.72$\pm$2.40 &  407.52$\pm$1.20 & 5003.83$\pm$0.10 &  -228.43$\pm$7.19 & 1013.86$\pm$3.59 & 39.67 & 39.65 & 38.15 \\ 
J1358+2658  & 42.36 & 5007.86$\pm$0.05 &   60.89$\pm$2.99 &  448.41$\pm$3.59 & 5003.64$\pm$0.24 &  -252.57$\pm$16.17 & 1366.38$\pm$5.39 & 40.05 & 39.99 & 39.16 \\ 
J1402+2159  & 41.49 & 5006.18$\pm$0.05 & - 39.70$\pm$2.99 &  291.89$\pm$1.80 & 5003.45$\pm$0.13 &  -163.17$\pm$8.98 &  762.86$\pm$5.39 & 38.34 & 38.24 & 37.67 \\ 
J1421+2824 & N.D. &    &            &         &                         &         &              & 42.20 & 42.18 & 40.66 \\ 
J1443+4725  & 42.32 & 4999.40$\pm$0.54 & -445.66$\pm$32.33 & 1199.99$\pm$33.53 &    0.00 &          0.00 &         0.00 & 42.98 & 42.97 & 41.43 \\ 
J1505+0326  & 41.77 & 5000.96$\pm$0.99 & -352.25$\pm$59.28 &  597.88$\pm$60.48 & 4995.52$\pm$4.34 &  -326.51$\pm$317.35 & 1180.25$\pm$131.13 & 43.09 & 43.09 &  0.00 \\ 
J1536+5433  & 41.06 & 5007.23$\pm$0.75 &   23.17$\pm$44.91 &  455.46$\pm$16.17 &         0.00 &    0.00 &         0.00 & 37.98 & 37.92 & 37.13 \\ 
J1537+4942  & 41.67 & 5006.48$\pm$0.08 &  -21.74$\pm$4.79 &  153.70$\pm$6.59 & 5000.77$\pm$0.37 &  -341.82$\pm$25.15 &  837.60$\pm$16.17 & 40.09 & 40.09 &  0.00 \\ 
J1548+3511  & 42.40 & 5006.63$\pm$0.07 &  -12.75$\pm$4.19 &  265.10$\pm$2.40 & 4999.95$\pm$0.59 &  -400.03$\pm$37.72 &  971.55$\pm$27.54 & 37.85 & 37.23 & 37.73 \\ 
J1555+1911  & 40.82 & 5005.91$\pm$0.43 &  -55.86$\pm$25.75 &  349.70$\pm$4.79 &    0.00 &       0.00 &         0.00 & 38.19 & 37.75 & 38.00 \\ 
J1559+3501  & 40.58 & 5005.48$\pm$0.07 &  -81.61$\pm$4.19 &  293.30$\pm$3.59 & 4999.65$\pm$0.38 &  -349.15$\pm$25.15 &  806.57$\pm$14.97 & 42.46 & 42.46 &  0.00 \\ 
J1612+4219  & 42.31 & 5004.23$\pm$0.14 & -156.46$\pm$8.38 &  741.71$\pm$12.57 & 5005.96$\pm$0.36 &   103.66$\pm$28.14 & 1913.50$\pm$20.96 & 39.86 & 39.45 & 39.65 \\ 
J1629+4007  & 42.06 & 5009.08$\pm$0.06 &  133.94$\pm$3.59 &  211.51$\pm$2.40 & 5004.52$\pm$0.34 &  -272.60$\pm$22.15 &  830.54$\pm$19.16 & 41.71 & 41.70 & 39.06 \\ 
J1633+4718  & 41.65 & 5006.57$\pm$0.05 &  -16.35$\pm$2.99 &  255.23$\pm$1.80 & 5001.91$\pm$0.23 &  -278.77$\pm$14.97 & 1185.89$\pm$13.17 & 40.67 & 40.64 & 39.46 \\ 
J1634+4809  & 41.96 & 5002.47$\pm$0.44 & -261.84$\pm$26.35 &  389.19$\pm$24.55 & 4993.44$\pm$1.69 &  -541.39$\pm$125.14 & 1109.74$\pm$56.28 & 41.42 & 41.40 & 40.03 \\ 
J1703+4540  & 41.55 & 5010.34$\pm$1.23 &  209.39$\pm$73.65 &  335.60$\pm$49.10 & 5004.73$\pm$3.96 &  -335.68$\pm$309.56 &  645.82$\pm$110.77 & 40.20 & 40.16 & 39.04 \\ 
J1709+2348  & 42.14 & 5008.43$\pm$0.09 &   95.02$\pm$5.39 &  401.88$\pm$3.59 & 5003.06$\pm$0.39 &  -321.53$\pm$26.94 &  786.83$\pm$9.58 & 39.55 & 39.55 &  0.00 \\ 
J1713+3523  & 40.98 & 4995.59$\pm$0.88 & -673.79$\pm$52.69 & 1641.35$\pm$47.90 &    0.00 &         0.00 &         0.00 & 39.54 & 39.51 & 38.33 \\ 
J2242+2943  & 41.34 & 5006.77$\pm$0.01 &   -4.46$\pm$0.72 &  226.27$\pm$3.75 & 4997.42$\pm$1.64 &  -564.31$\pm$98.63 & 1089.89$\pm$279.06 & 38.88 & 38.62 & 38.54 \\ 
J2314+2243  & 42.31 & 5006.86$\pm$0.81 &    1.02$\pm$48.50 &  602.11$\pm$25.75 & 4989.64$\pm$2.20 & -1031.42$\pm$178.43 & 1579.30$\pm$55.69 & 40.48 & 40.42 & 39.58 \\ 
J1029+5556  & 41.82 & 5006.96$\pm$0.22 &    6.97$\pm$13.39 &  335.37$\pm$37.37 & 5004.16$\pm$1.65 &  -160.73$\pm$98.89 & 2004.90$\pm$387.54 & 38.46 & 38.46 &  0.00 \\ 
J1228+5017  & 41.83 & 5008.56$\pm$0.03 &  102.76$\pm$1.71 &  218.21$\pm$5.78 & 5003.58$\pm$0.27 &  -195.21$\pm$16.48 &  978.24$\pm$47.52 & 39.00 & 38.91 & 38.27 \\ 
J1232+4957  & 41.66 & 5003.32$\pm$1.28 & -211.09$\pm$76.44 &  578.97$\pm$156.31 & 4994.82$\pm$4.45 &  -719.77$\pm$266.80 & 1142.04$\pm$297.77 & 38.23 & 38.23 &  0.00 \\ 
J1509+6137  & 40.94 & 5003.87$\pm$0.12 & -178.29$\pm$7.34 &  320.40$\pm$24.24 & 5005.87$\pm$1.91 &   -58.51$\pm$114.53 & 1690.48$\pm$189.36 & 37.68 & 37.68 &  0.00 \\ 
J1510+5547  & 41.29 & 5006.85$\pm$0.03 &    0.67$\pm$1.56 &  123.77$\pm$8.75 & 5004.09$\pm$0.29 &  -164.99$\pm$17.14 &  547.56$\pm$34.91 & 37.69 & 37.48 & 37.27 \\ 
J1522+3934  & 40.30 & 5007.43$\pm$0.36 &   35.10$\pm$21.3 &  321.12$\pm$79.06 & 5001.67$\pm$3.29 &  -310.03$\pm$197.07 & 1306.38$\pm$324.23 & 37.95 & 37.88 & 37.12 \\ 
J1641+3454  & 41.45 & 5006.61$\pm$0.07 &  -14.01$\pm$4.46 &  448.44$\pm$19.38 & 5006.19$\pm$0.62 &   -39.20$\pm$37.30 & 1274.45$\pm$136.69 & 38.92 & 38.79 & 38.33 \\ 
\hline
\end{longtable}
\end{landscape}

\end{document}